\documentclass[sigplan,nonacm]{acmart}

\usepackage[]{hyperref}
\usepackage{algorithmic}
\usepackage{graphicx}
\usepackage{textcomp}
\usepackage{xcolor}
\usepackage{array}
\usepackage{makecell}
\usepackage{diagbox}

\usepackage[normalem]{ulem}
\usepackage{listings}
\usepackage{subfig}
\usepackage{multirow}
\usepackage{multicol}
\usepackage{booktabs} 
\usepackage[ruled,vlined,linesnumbered]{algorithm2e}
\usepackage[most]{tcolorbox}
\usepackage{cleveref}
\usepackage{lipsum}

\lstdefinestyle{code_style}{
  language=bash,
  basicstyle=\footnotesize\ttfamily,
  numbers=none,
  keywordstyle=\color{black},
  numberstyle=\tiny,
  numbersep=4pt,
  frame=tblr,
  columns=fullflexible,
  backgroundcolor=\color{lightgray!20},
  linewidth=0.95\linewidth,
  xleftmargin=0.03\linewidth
}
\makeatletter
\newcommand{\algorithmfootnote}[2][\small]{%
  \let\old@algocf@finish\@algocf@finish
  \def\@algocf@finish{\old@algocf@finish
    \leavevmode\rlap{\begin{minipage}{\linewidth}
    #1#2
    \end{minipage}}%
  }%
}

\newif\ifhighlight
\highlightfalse

\newenvironment{CompactItemize}%
{\begin{list}{$\bullet$}%
    {\leftmargin=\parindent \itemsep=2pt \topsep=2pt
    \parsep=0pt \partopsep=0pt}}%
{\end{list}}

\renewcommand\footnotetextcopyrightpermission[1]{}
\def\projectname{Aegis}

\begin{document}

\title{\projectname{}: Taxonomy and Optimizations for Overcoming Agent-Environment Failures in LLM Agents}
\author{Kevin Song}
\affiliation{%
  \institution{University of Toronto, Vector Institute}
  \city{Toronto}
  \country{Canada}
}
\email{xinyang.song@utoronto.ca}

\author{Anand Jayarajan}
\affiliation{%
  \institution{University of Toronto, Vector Institute}
  \city{Toronto}
  \country{Canada}
}
\email{anandj@cs.toronto.edu}

\author{Yaoyao Ding}
\affiliation{%
  \institution{University of Toronto, Vector Institute}
  \city{Toronto}
  \country{Canada}
}
\email{yaoyao@cs.toronto.edu}

\author{Qidong Su}
\affiliation{%
  \institution{University of Toronto, Vector Institute}
  \city{Toronto}
  \country{Canada}
}
\email{qdsu@cs.toronto.edu}

\author{Zhanda Zhu}
\affiliation{%
  \institution{University of Toronto, Vector Institute}
  \city{Toronto}
  \country{Canada}
}
\email{zhandazhu@gmail.com}

\author{Sihang Liu}
\affiliation{%
  \institution{University of Waterloo}
  \city{Waterloo}
  \country{Canada}
}
\email{sihangliu@uwaterloo.ca}

\author{Gennady Pekhimenko}
\affiliation{%
  \institution{University of Toronto, Vector Institute}
  \city{Toronto}
  \country{Canada}
}
\email{pekhimenko@cs.toronto.edu}



\begin{abstract}
Large Language Models (LLMs) agents augmented with domain tools promise to autonomously execute complex tasks requiring human-level intelligence, such as customer service and digital assistance. 
However, their practical deployment is often limited by their low success rates under complex real-world environments.
To tackle this, prior research has primarily focused on improving the agents themselves, such as developing strong agentic LLMs, while overlooking the role of the system environment in which the agent operates.

In this paper, we study a complementary direction: improving agent success rates by optimizing the system environment in which the agent operates.
We collect 142 agent traces (3,656 turns of agent-environment interactions) across 5 state-of-the-art agentic benchmarks. 
By analyzing these agent failures, we propose a taxonomy for agent-environment interaction failures that includes 6 failure modes.
Guided by these findings, we design \projectname{}, a set of targeted environment optimizations: 1) environment observability enhancement, 2) common computation offloading, and 3) speculative agentic actions.
These techniques improve agent success rates on average by $6.7-12.5\%$, without any modifications to the agent and underlying LLM. 

\end{abstract}

\maketitle 
\pagestyle{plain} 

\section{Introduction}

Large Language Models (LLMs) are increasingly being deployed as autonomous agents to perform complex, multi-turn real-world tasks.
LLM agents are equipped with tools, usually in the form of functions, to interact with the outside world, such as database queries \cite{spider1, spider2, text2sql_survey}, file system operations \cite{bfcl, sweagent, agent_fs}, and web searches \cite{agentoccam, webarena, chatgpt_agent}.
The combination of powerful foundation models and domain-specific tools enables autonomous systems that can perform complex real-world tasks, such as customer service \cite{taubench, proxyllm}, business data management \cite{crmarena, crmarena_pro}, and digital assistant \cite{llm_assist, personal_llm}.  

Despite this promise, practical agent deployments suffer from low task success rates \cite{crmarena_pro, llms_lost, llm_distraction, agent_misalign, agent_safe, andriod_arena}.
Agent failures occur when agents perform tasks incorrectly, producing misleading output or undesirable side effects, requiring costly human interventions \cite{agent_fail_cost1, agent_fail_cost2, agent_fail_cost3}.
To address this challenge, prior works focus on improving the agent/LLM, such as building foundation LLMs with stronger agentic abilities \cite{o3, gemini25} or fine-tuning LLMs for specific workloads \cite{xlam, xlam3, xlam2}.
We refer to this paradigm as \textit{agent-for-system}. 
However, while state-of-the-art LLMs demonstrate strong intrinsic abilities in mathematics and logic \cite{o1, o3, dsr1, gemini_deep_think}, their performance on real-world agentic tasks remains limited.
OpenAI o3 \cite{o3}, a model optimized for agentic use cases, improves success rates on customer service workloads \cite{taubench} by only 2-3\%  \cite{o3} over its predecessor o1, illustrating that strengthening models alone has limited effectiveness.

To understand why agents fail despite strong underlying LLMs, we conducted a preliminary analysis of failure traces from state-of-the-art agentic benchmarks \cite{bfcl, taubench}. 
We found that failures stem not only from flawed LLM reasoning, but also from the agent's interaction with its surrounding environment, such as inability to discover information, follow workload constraints, and process large volumes of data \cite{spider2, crmarena_pro, why_multiagent_fail}. 
This led us to ask: \textit{How significant is the system environment for agent reliability, and how much improvement can environment optimizations achieve?}
In this paper, we investigate this complementary but largely overlooked direction: optimizing the system environment to improve agent reliability, an approach which we term \textit{system-for-agent.}

\begin{table*}[t]
\small
\centering
\caption{Taxonomy of agent system-interaction failures and corresponding environment optimizations.}
\label{tab:failure_taxonomy}
\begin{tabular}{@{} p{2.2cm} p{3cm} p{6.5cm} p{5cm} @{}}
\toprule
\textbf{Category} & \textbf{Subcategory} & \textbf{Description} & \textbf{Optimization} \\
\midrule
\multirow{5}{*}{\parbox{2cm}{\raggedright Exploration Failures (\autoref{subsec:exploration_failures})}}
  & State-space Navigation Failure & Agent fails to navigate the environment to retrieve all necessary data required to complete the task. & Environment lookahead (\autoref{subsec:opt1}) \\
\cmidrule(lr){2-4}
  & State Awareness Failure & Agent has incorrect understanding about its current position within the environment. & Explicit agent state changes (\autoref{subsec:opt1}) \\
\midrule
\multirow{7}{*}{\parbox{2cm}{\raggedright Exploitation Failures (\autoref{subsec:exploitation_failures})}}
  & Tool Output Processing Failure & Agent makes computational errors (e.g., comparisons, ranking) when processing information gathered from tool outputs. & Offload common computations (\autoref{subsec:opt2}) \\
\cmidrule(lr){2-4}
  & Domain Rule Violation & Agent fails to follow domain rule by either performs forbidden actions or incorrectly blocks valid actions. & Offload domain rule validations (\autoref{subsec:opt2}) \\
\cmidrule(lr){2-4}
  & User Instruction Following Failure & Agent fails to follow user's specific instructions as requested. & No direct environment optimizations \\
\midrule
\multirow{2}{*}{\parbox{2.5cm}{\raggedright Resource Exhaustion (\autoref{subsec:turn_limit_failures})}}
  & -- & Agent fails due to exceeding allocated maximum number of turns or tokens before task completion. & Speculative agentic actions (\autoref{subsec:opt3}) \\
\bottomrule
\end{tabular}
\end{table*}

However, designing effective environments for agents is non-trivial, as real-world system environments are complex and vary widely across different domains.
To bridge this gap, we present an in-depth study of agent-environment interactions and a systematic methodology for optimizing system environments to improve agent reliability.
We collect agent failure traces comprising 142 failed agent tasks and 3,656 turns of agent interactions.
To precisely localize agent failures, we introduce a \textit{subtask-based abstraction} inspired by Hierarchical Task Networks (HTNs) from the automated planning literature \cite{htn, htn2}.
%
%
Using this abstraction, we annotate the collected failure traces to identify the specific subtask where each failure first occurred.
Based on these annotations, we propose a taxonomy of agent-environment interaction failures, summarized in Table \ref{tab:failure_taxonomy}.
We categorize failures into three main types: 1) \textit{Exploration failures}, which occur when agent fails to gather all information required to complete the task, 2) \textit{Exploitation failures}, which arise when agent incorrectly processes information it has collected, and 3) \textit{Resource exhaustion}, which accounts for failures due to exceeding the pre-allocated turns/token limit.

Based on our analysis, we propose \projectname{}, a set of system-level optimizations to address the identified failure modes, as summarized in Table \ref{tab:failure_taxonomy}. 
To mitigate exploration failures, we enhance environment observability through expanding the agent's observation window and explicitly communicating state changes. 
To reduce exploitation failures, we offload deterministic reasoning operations such as ranking and rule-checking from the agent to be performed in the environment instead. 
Lastly, we introduce speculative agentic actions to address resource exhaustion by preemptively bundling related tool calls, reducing turn count and token consumption.

We evaluate the effectiveness of our environment optimizations across 5 state-of-the-art agent benchmarks, achieving $6.7-12.5\%$ average success rate improvements on average without any modifications to the agent. 
For context, this improvement is comparable to or greater than those typically seen between major LLM model generations, which are typically a few percent \cite{o3, claude4, gpt41}.
We further validate these improvements using a model fine-tuned for agentic workloads: xLAM-2 8B~\cite{xlam3}.
Our environment optimizations increase success rates by $3.3-8.8\%$ on average, demonstrating that system-for-agent complements model fine-tuning. 
Lastly, we observe an unexpected benefit of environment optimizations: the monetary cost of LLM inference API is reduced by $7.1-17.7\%$ as a result of more efficient agent-environment interactions.



In summary, we make the following contributions:
\begin{CompactItemize}
  \item We collect and analyze 142 failed tasks and 3,656 agent interactions across 5 diverse agentic workloads. Based on this analysis, we propose a taxonomy for categorizing agent-environment interaction failures, consisting of 6 distinct failure modes.
  \item We introduce \projectname{}, a set of targeted environment optimizations to address each type of failure in our taxonomy.
  \item We implement and evaluate system environment optimizations on agentic benchmarks, demonstrating 6.7-12.5\% improvements in task success rates on 3 representative LLM models, GPT 4.1, GPT 4.1 mini, and o3, without modifying the agent or underlying LLM model.
\end{CompactItemize}

\section{Background} \label{sec:background}

\begin{table*}[h!]
\small
\centering
\caption{Agent goals, environments, and key tools for each workload.}
\label{tab:agent_workloads}
\begin{tabular}{
>{\raggedleft\arraybackslash}p{2.0cm}
>{\raggedright\arraybackslash}m{3.8cm}
>{\raggedright\arraybackslash}m{5.0cm}
>{\raggedright\arraybackslash}m{5.5cm}
}
\toprule
\textbf{Domain} & \textbf{Agent Goal} & \textbf{Environment} & \textbf{Key Tools} \\
\midrule
Airline \cite{taubench} & Book or modify flights & 3 DBs for user, reservations, and flights & \texttt{get\_user\_details}, \texttt{search\_direct\_flight}, \texttt{modify\_flight} \\
\midrule
Retail \cite{taubench} & Return or exchange orders & 3 DBs for user, orders, and products & \texttt{get\_user\_details}, \texttt{get\_order\_details}, \texttt{return\_item} \\
\midrule
File system \cite{bfcl} & File system operations & Emulated Linux file system & \texttt{cd}, \texttt{ls}, \texttt{rm}, \texttt{cp} \\
\midrule
CRM  \cite{crmarena} & Customer Relationship Management (CRM) tasks & Realistic Salesforce platform (SQL DBs) & \texttt{get\_cases}, \texttt{calculate\_avg}, \texttt{get\_shipping\_state} \\
\midrule
Medical \cite{medagentbench} & Mmanaging medical records & Medical records DB over HTTP & \texttt{GET Patient}, \texttt{GET Record} \\
\bottomrule
\end{tabular}
\end{table*}

\subsection{Large Language Model Agents}
\begin{figure}
  \centering
  {\includegraphics[width=0.8\columnwidth]{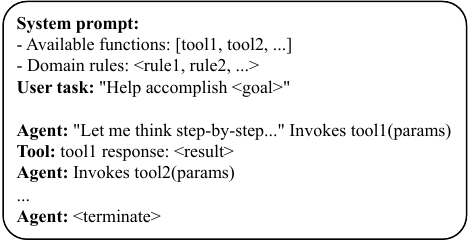}}
  \caption{Typical agent execution flow.}
\label{fig:agent_bench}
\Description{}
\end{figure}

Large Language Model (LLM)-based agents augment LLMs with the ability to call external tools or functions, enabling them to complete complex tasks autonomously. Through tool use, agents can gather information and manipulate their surrounding system environment to complete a variety of objectives.
A typical agent execution trace is illustrated in Figure~\ref{fig:agent_bench}. The agent is first presented with a system prompt that specifies the set of available functions, their arguments, descriptions, and return values. The system prompt may also define domain rules that constrain the agent’s behavior, such as prohibiting certain actions.
The agent is given a task, such as retrieving data, performing analysis, or modifying the environment.
During execution, the agent proceeds in discrete steps. In each step, the agent may reason and optionally invoke tools via a function calls, and the environment returns a response.
This loop continues until the agent decides to terminate and optionally returns a final answer.

To evaluate agent performance on real-world tasks, several benchmarks have been proposed. In this work, we study five state-of-the-art benchmarks (Table~\ref{tab:agent_workloads}), each of which includes an environment, available tools, and a set of tasks. For each task, the benchmark defines an expected execution, including the correct sequence of tool calls and the final environment state. Agent correctness is measured by comparing its actual trace against these expectations (e.g., whether the correct tools were invoked in the right order, and whether the final environment state matches the expected outcome).

\begin{figure}
  \centering
  {\includegraphics[width=\columnwidth]{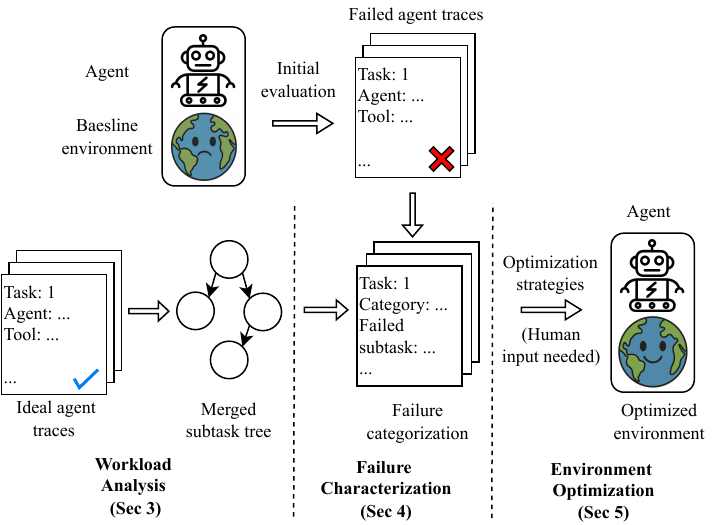}}
  \caption{Methodology overview.}
\label{fig:flow}
\Description{}
\end{figure}

\subsection{The Accuracy Challenge of LLM Agents}
Despite their promise, LLM agents often fail to complete complex, real-world tasks~\cite{agent_fail_cost2, why_multiagent_fail, crmarena_pro, which_agent_causes_failures}. 
This is because real-world workloads demand specialized domain knowledge, interaction with complex environments, and processing of large-scale data~\cite{spider2, llm_distraction, crmarena_pro, why_multiagent_fail, agentbench}. 
In practice, failures require costly fallback mechanisms: humans may need to take over incorrectly performed tasks, or the agent must re-execute the task from scratch, both of which reduce the efficiency gains promised by automation \cite{agent_fail_cost1, agent_fail_cost2, agent_fail_cost3, agent_fail_cost4, agent_long_tasks}.

Current efforts to improve agent reliability primarily focus on enhancing the underlying LLM. One strategy is to improve the general capabilities of foundation LLM models. 
For instance, state-of-the-art LLMs such as GPT-4.1 and o3 are explicitly optimized for agentic use cases \cite{gpt41, o3}.
Another approach is to fine-tune LLM for a specific workload. This requires building custom pipelines for data collection, model training, and evaluation \cite{xlam, xlam2, xlam3, toolace, wattool}. The Salesforce xLAM-2 models, for example, are fine-tuned from Llama base models and demonstrated a higher task success rate than their base counterparts \cite{xlam}.
While these agent-for-system strategies are effective at improving agent capabilities, they require a costly and complex multi-stage pipeline of data preparation, model
training, and evaluation \cite{xlam, xlam2, xlam3}.

In this work, we study a complementary but overlooked direction: enhancing agent reliability by optimizing its surrounding environment. 
Our methodology, illustrated in \autoref{fig:flow}, consists of three steps. 
Workload Analysis (\autoref{sec:workload_analysis}): we analyze the fundamental structure of each workload using a subtask abstraction to establish a baseline for correct agent behavior. 
Failure Taxonomy (\autoref{sec:agent_failures}): captures agent failures stemming from environment interactions and identifies recurring failure patterns. 
Environment Optimizations (\autoref{sec:opts}): target common failure modes to improve reliability.

\section{Agentic Workload Analysis} \label{sec:workload_analysis}

In this section, we study properties of agentic workloads and describe the subtask abstraction for agent-environment interaction analysis. 

\begin{figure*}[t]
  \centering
{\includegraphics[width=1\linewidth]{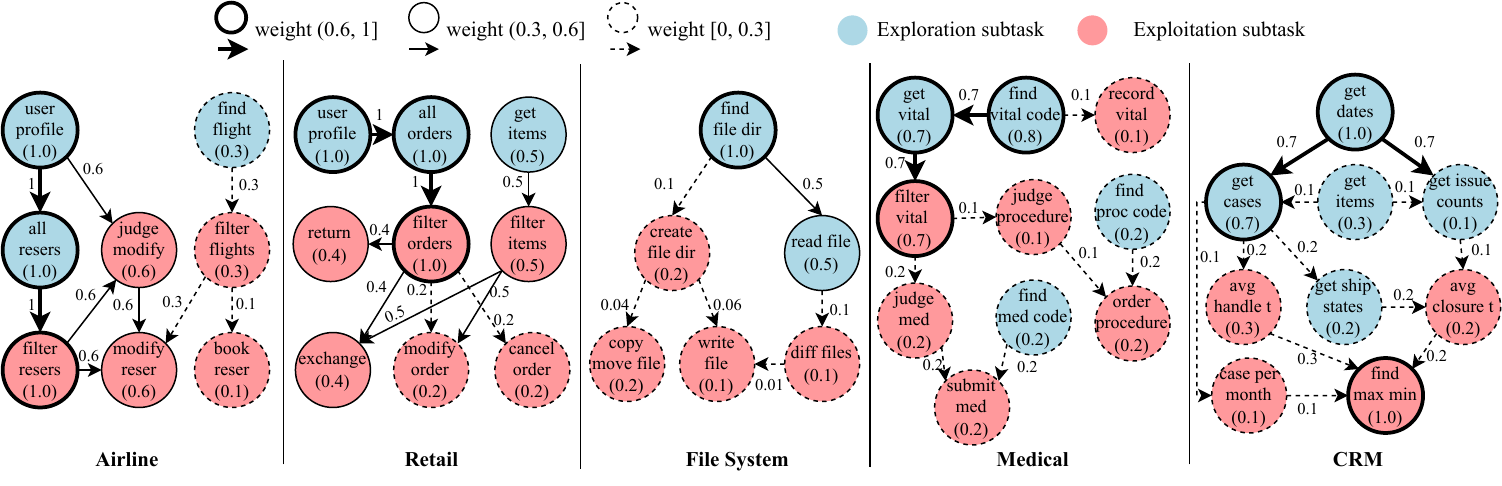}}
  \caption{Merged subtask graphs. Node/edge thickness represent weight. Color encodes explore/exploit subtask.}
\label{fig:merged_trees}
\Description{}
\end{figure*}

\begin{figure}
  \centering
  \subfloat[Node weight distributions.]{
  \includegraphics[width=\columnwidth]{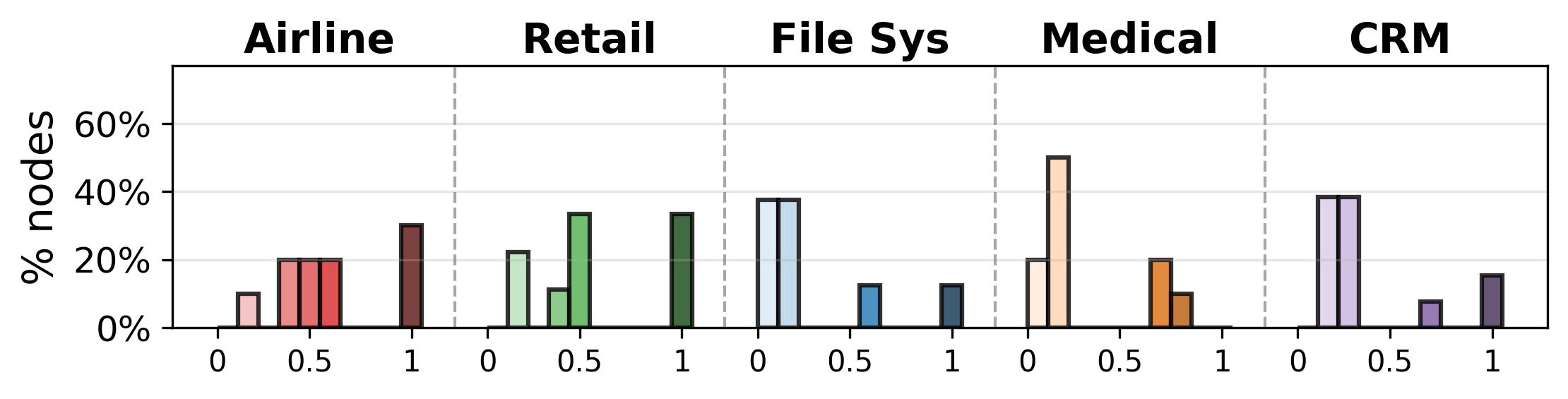}}
  \hspace{0.3mm}
  \subfloat[Edge weight distribution.]{
  \includegraphics[width=\columnwidth]{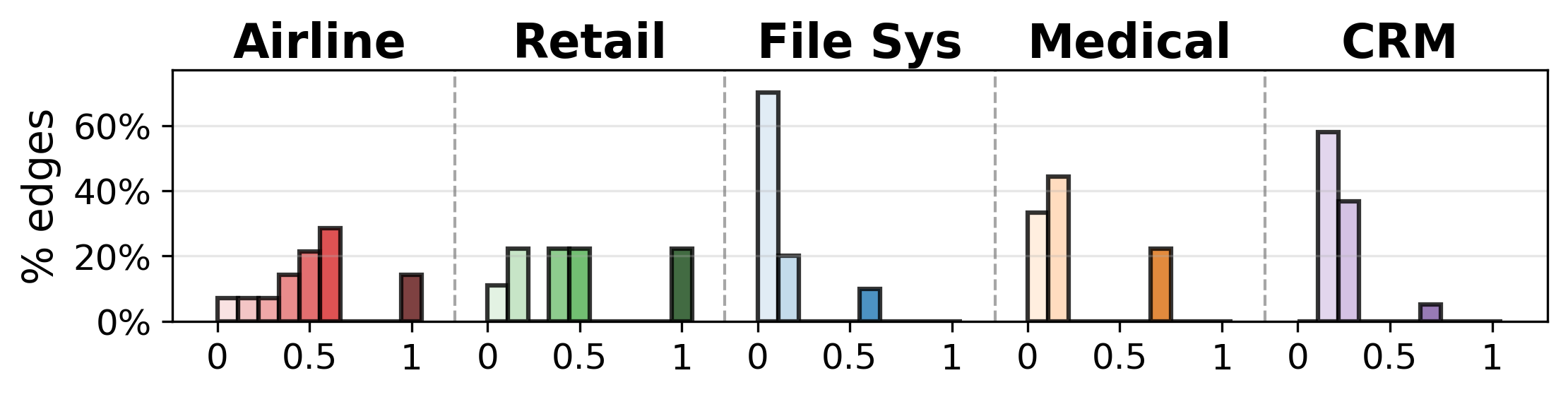}}
  \caption{Distribution of node and edge weights in each merged subtask graph. X-axis shows weight ranges for each workload from 0 to 1. Y-axis shows percent of nodes/edges that fall within each bucket.}
  \Description{}
  \label{fig:subtask_weight_distribution}
\end{figure}

\subsection{Methodology}
We first establish the following definitions:

\begin{CompactItemize}
    \item Agent: LLM operating as an autonomous decision-maker that initiates actions to complete user objectives\footnote{In this work, we focus on single-agent systems, though our taxonomy and principles could extend to multi-agent systems in future work}.
    
    \item System environment: The surrounding system with which an agent interacts, such as databases and file systems. 
    
    \item Task: A problem assigned to an agent. In this paper, each task corresponds to one entry in an agent benchmark.
    
    \item Agent trace: A sequence of actions performed by an agent to fulfill a task, successfully or unsuccessfully. 
\end{CompactItemize}

We collect agent traces from the five benchmarks detailed in \autoref{tab:agent_workloads}. From each benchmark, we randomly sample 50 tasks, for a total of 250 tasks. We partition the tasks in each workload into an \textit{analysis set} (30 tasks) and an \textit{evaluation set} (20 tasks). We perform failure analysis and develop environment optimizations on the analysis set (\autoref{sec:workload_analysis} and \autoref{sec:agent_failures}) and then measure agent success rates on the unseen evaluation set to test the generalizability of our optimizations.

\subsection{The Subtask Abstraction} \label{sec:subtask_abstraction}
We adapt the Hierarchical Task Network (HTN) framework from automated planning literature \cite{htn}. HTN decomposes complex tasks into manageable components with explicit dependencies. We use two key HTN concepts:

\begin{CompactItemize}
    \item Subtask: A discrete, logical unit of work that contributes to the completion of the overall task.
    \item Subtask graph: A partial ordering of subtasks, where nodes represent subtasks and edges represent dependencies between subtasks.
\end{CompactItemize}

Consider an airline workload task: ``Change my flight tomorrow to the cheapest option''. This decomposes into four subtasks: 1) retrieve user profile, 2) get all user reservations, 3) find the cheapest alternative flight, and 4) modify the reservation. The dependencies are: subtask 4 depends on subtasks 2 and 3 (the agent needs both the correct reservation and cheapest flight), while subtask 2 depends on subtask 1 (user profile is required to retrieve reservations).

The key benefit of the subtask abstraction is that it enables us to localize failures to the precise point agent deviates from the correct solution. 
Prior agent failure studies \cite{why_multiagent_fail, trail, which_agent_causes_failures} adopt a task-level failure attribution method, which assigns one failure label, such as "Incorrect Information Retrieval", for each failed agent task.
We find this method insufficient for analyzing agent-environment failures, as it does not reveal which part of the system environment should be optimized. 
For instance, "Incorrect Information Retrieval" failures can be due to different underlying reasons, such as misinterpreting a specific tool response, failing to retrieve a particular piece of data, or misunderstanding the task.

We derive subtasks in two steps. 
First, each tool call is mapped to a subtask. In the airline example, this yields 4 subtasks: \texttt{get\_user},  \texttt{get\_reservations}, \texttt{find\_flights}, and \texttt{modify\_reservation}. 
Second, subtasks that encompass multiple logical operations are split. 
For instance, we decompose \texttt{modify\_reservation} into \texttt{judge modify} (reasoning about modification eligibility according to airline domain rules) and \texttt{modify reservation} (executing the change). 
This decomposition is critical because it allows us to distinguish between different failure types, which require different solutions.
In this case, a \texttt{modify\_reservation} could stem from two distinct causes: an error in judging the modification's eligibility or an error in executing the change itself. 

\subsection{Exploration and Exploitation Subtasks}
We divide subtasks into two categories: \textit{exploration} and \textit{exploitation}. 
Exploration subtasks gather new information from the environment (e.g., listing flights on a given date, retrieving patient records, browsing directory contents). 
Exploitation subtasks act on knowledge the agent already possesses (e.g., changing a reservation, deleting a file, ordering a medical procedure).
We make this distinction as exploration and exploitation subtasks benefit from distinct environment optimizations. 
Exploration subtasks require efficient and accurate information collection, benefiting from environments with high observability. 
Exploitation subtasks require environments that reduce the likelihood of action errors when the agent processes complex information.
This distinction drives our optimization strategy. 
Section \ref{sec:opts} demonstrates how we design targeted environment improvements for exploration and exploitation failures.

\subsection{Workload Analysis Insights} \label{subsec:workload_insights}
To analyze agent workload properties, we construct \emph{merged subtask graphs} by first building individual subtask graphs for each task in each workload (30 tasks in the analysis set), then aggregating them into a single merged graph.
The nodes and edges of the merged graph are the union of those in the individual graphs. 
A node's weight is the fraction of tasks in which the subtask appears, while an edge's weight is the fraction of tasks where the corresponding dependency exists. 
For example, a node appearing in 3 of 30 tasks has a weight of 0.1, and an edge appearing in 15 of 30 tasks has a weight of 0.5.
Intuitively, node weights indicate how likely an ideal agent would perform each subtask, while edge weights indicate transition probabilities between subtasks.

Figure~\ref{fig:merged_trees} shows the merged subtask graph for each workload, and Figure~\ref{fig:subtask_weight_distribution} plots the weight distributions of nodes and edges. Our analysis reveals two key properties.
First, \textbf{agent workloads contain frequently recurring subtasks}, represented by high-weight nodes. As shown in Figure \ref{fig:subtask_weight_distribution}a, all workloads except the file system have subtasks that appear in over 70\% of tasks. These common subtasks typically occur early in the workflow and involve \emph{retrieving critical data objects} required by subsequent steps. For instance, in the airline workload, an agent must retrieve user and reservation details before it can modify a booking. 
%
This observation motivates environment optimizations that improve the reliability of common subtasks, as detailed in Section \ref{sec:opts}.
Second, \textbf{workloads exhibit common subtask transitions}, indicated by high-weight edges (Figure~\ref{fig:subtask_weight_distribution}b). A high weight edge from subtask A to B means that performing A is very likely to be followed by performing B. For instance, in the retail workload, retrieving a user's profile is followed by retrieving their order history in all tasks. 
This motivates speculative agentic action optimization, as detailed in Section \ref{sec:opts}.

\section{Failure Taxonomy}  \label{sec:agent_failures}
\begin{figure}
  \centering
  \subfloat[GPT 4.1]{
  \includegraphics[width=\columnwidth]{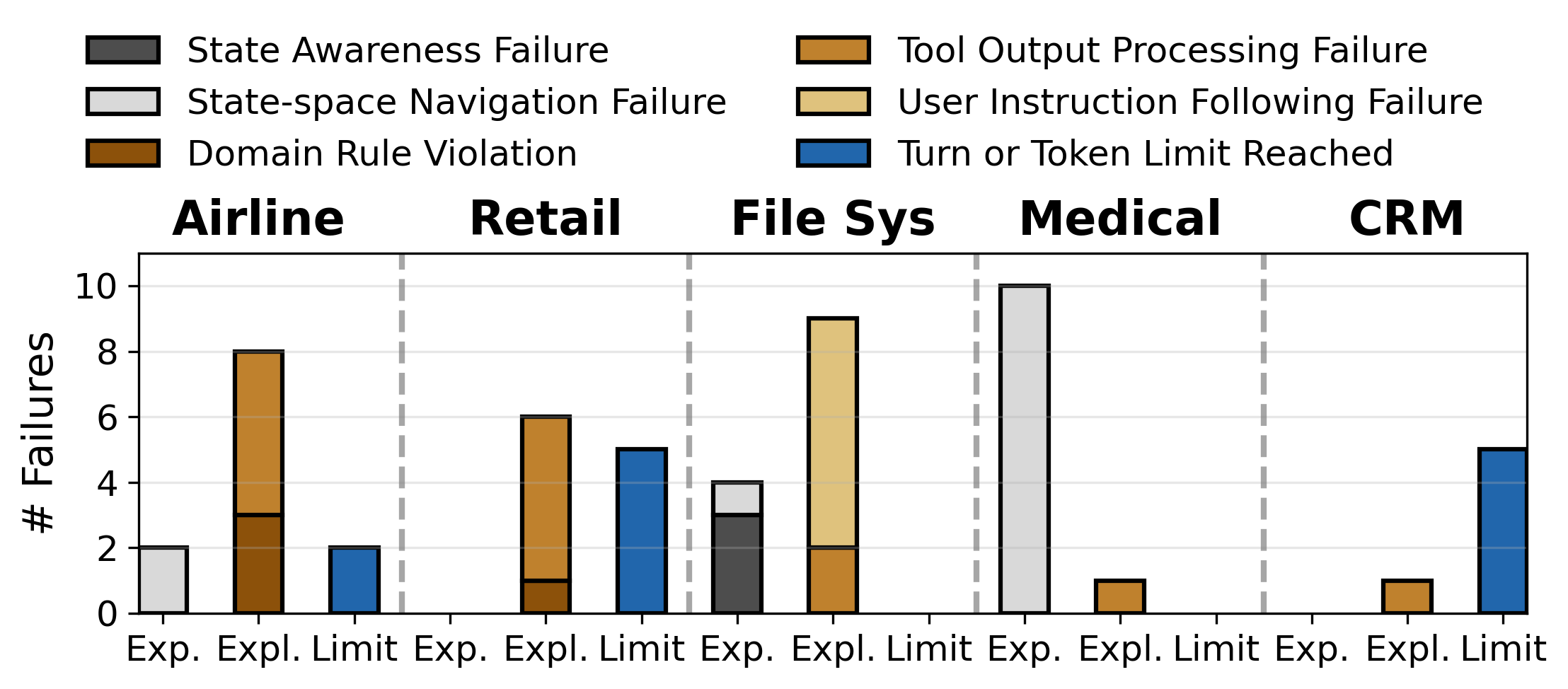}}
  \hspace{0.3mm}
  \subfloat[GPT 4.1 mini]{
  \includegraphics[width=\columnwidth]{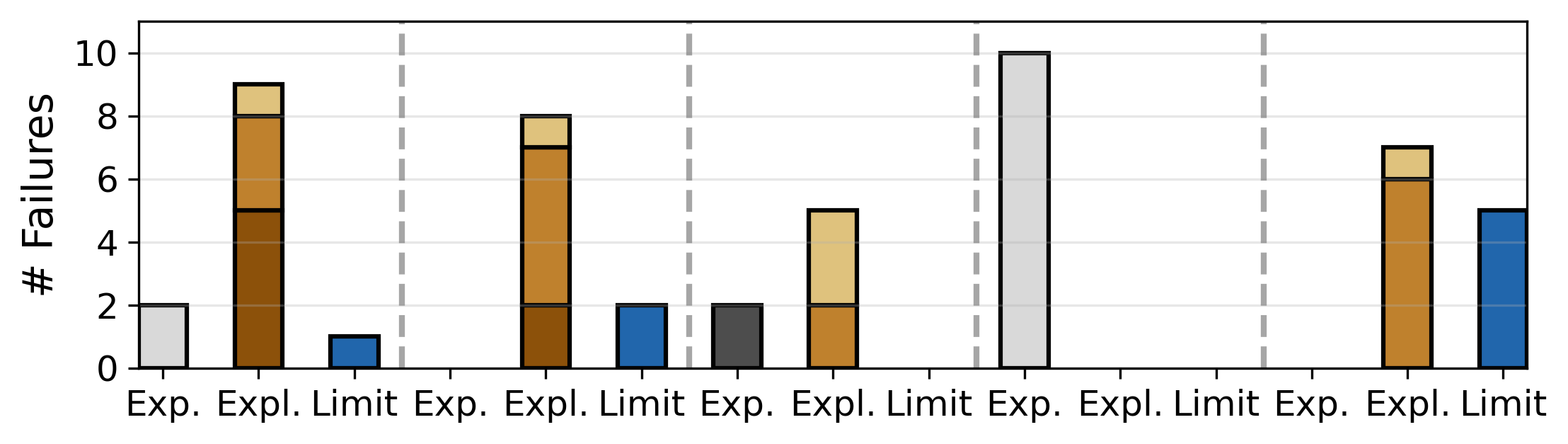}}
    \hspace{0.3mm}
\subfloat[o3]{
  \includegraphics[width=\columnwidth]{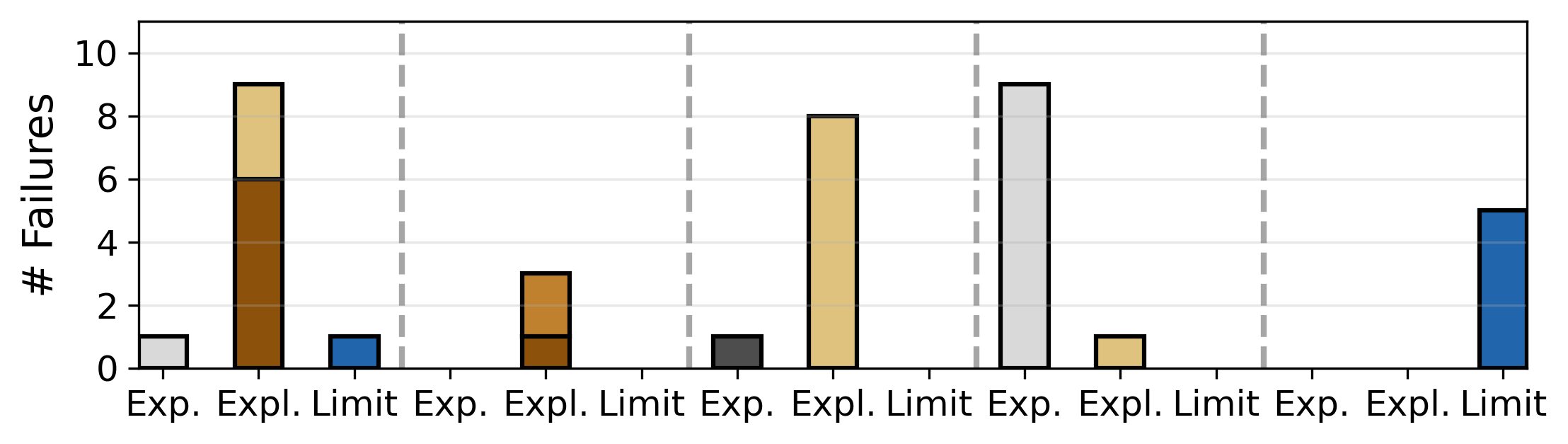}}
  \caption{Failure category distribution. Exp: exploration. Expl: exploitation. Limit: resource exhaustion.}
  \Description{}
  \label{fig:failure_distr}
\end{figure}

This section presents our taxonomy of agent environment interaction failures.
We define a \textit{failure} as the first unsuccessful subtask (as defined in Section~\ref{sec:subtask_abstraction}) in an agent's execution trace. 
This focus on the first failure allows us to identify the root cause of errors, including cascading failures where an early mistake (e.g., retrieving the wrong data) makes all subsequent steps incorrect. 
We do not classify transient errors that the agent self-corrects without impacting the final outcome, such as a corrected command syntax, as failures.
Applying this methodology, we analyzed 142 failed agent traces across 5 workloads and 3 models, containing 3,656 turns of agent-environment interactions. 
For each of the failed traces, we annotate the subtask where the failure occurred and the failure category.
Table~\ref{tab:failure_taxonomy} presents our complete failure taxonomy, and Figure~\ref{fig:failure_distr} shows the distribution of failure categories across workloads and models.
We now explain each failure category in detail.

\subsection{Exploration Failures} \label{subsec:exploration_failures}
Exploration failures occur when an agent fails to gather information critical to completing its task. Exploration failures account for 30\%, 27\%, and 29\% of all failures under GPT 4.1, GPT 4.1 mini, and o3, respectively.
We conceptualize agent exploration as a search process where agents use tools to navigate an environment and collect information.
For instance, in a file system, the agent uses the \texttt{cd} tool to navigate to various directories to gather locations of different files.
The search process can be multi-step.
In the medical workload, to retrieve records associated with a specific patient, the agent must first invoke the \texttt{GET Patient} tool, then use information from the tool response to invoke \texttt{GET Record\&patient\_id}.
The goal of exploration is to find the \textit{critical context}: the minimal set of information required to complete the given task.
For instance, to manipulate a specific file, the agent must at least know the file's location within the file system.
Using this conceptualization of agent exploration, we identify two distinct subtypes of this failure:

\textbf{State-space Navigation Failure} occurs when the agent fails to navigate the environment to collect the critical context. 
This often happens when the agent's limited view of the environment causes incorrect navigation.
This is illustrated in Figure \ref{fig:explore_opt_illustrate}b, where the critical context lies beyond the agent's observation window. 
The agent then explores the environment in a best-effort manner, hoping to reach the critical information.
This failure mode is dominant in the medical workload, accounting for 91\%, 100\%, and 90\% of failures under GPT 4.1, GPT 4.1 mini, and o3, respectively.
A common type of failure occurs when a medical agent fails to retrieve a patient's complete medical history. The \texttt{GET Record} tool returns partial patient records by default for better resource efficiency. 
However, the agent is often not aware of this output truncation, causing it to perform downstream tasks based on this incomplete information.  

\textbf{State Awareness Failure} occurs when the agent loses track of its current position within the environment during navigation. This is distinct from a navigation failure, where the agent correctly identifies its position but does not know how to navigate to find information.
In this case, the agent's internal representation of its current state is incorrect. For example, in the file system workload, agents often misidentified their current working directory, leading them to execute file operations in the wrong location. This failure accounted for 23\% and 29\% of failures in the file system workload under GPT 4.1 and GPT 4.1 mini, respectively.
This type of failure is less common under o3, accounting for 10\% of all failures.

\subsection{Exploitation Failures} \label{subsec:exploitation_failures}
Exploitation failures occur when an agent possesses the necessary information gathered through exploration but fails to utilize this information correctly.
Exploitation failures account for the majority of failures in nearly all workloads (Figure~\ref{fig:failure_distr}), comprising 75\% of failures in airline, 88\% in retail, 60\% in file system, and 33\% in CRM under GPT 4.1. 
We categorize exploitation failures into three subtypes based on the source of the information that the agent failed to process correctly: tool outputs, domain rules, and user instructions.

\textbf{Tool Output Processing Failure} occurs when the agent incorrectly utilizes information gained from tool outputs in previous turns. 
Tool output processing failure is particularly common in the airline, retail, and CRM workloads, accounting for 25\%, 67\%, and 33\% of their respective failed tasks under GPT 4.1. 
For instance, a CRM agent correctly gathered all relevant customer service handling times but miscalculated the average case handling time.
We observe that this failure commonly arises from incorrect logical or mathematical operations performed by the agent. 
We further classify them based on the operation which the agent performs incorrectly: comparison (e.g., all items with price less/greater than threshold), calculation (e.g., total price), retrieval (e.g., find the most expensive item), and sorting (e.g., earliest/latest arrival time), which comprises 33\%, 25\%, 25\%, and 17\% of all tool output processing failures, respectively.

Surprisingly, we observe that agent fails at relatively simple operations such as comparisons. 
We confirm this by prompting the agent with only the essential information for the failed operation. For instance, we prompt the agent to find the average time from a list by providing only the list of time intervals. We find that in this case the LLM can consistently perform the previously failed operation correctly.
We conclude that the agent's failure to perform these operations is not an intrinsic limitation of the LLM, but a consequence of \textit{context distractions}~\cite{llm_distraction}. 
State-of-the-art LLMs are capable of solving difficult math problems~\cite{o1, o3, dsr1, gemini_deep_think}. 
However, in agentic workloads, the agent's context is crowded with additional information, such as system prompts, user instructions, and tool outputs that are not immediately relevant to the current operation.
As a result of the fundamental nature of the attention mechanism behind LLMs, their reasoning quality degrades under this distraction~\cite{llm_distraction, llms_lost}.

\textbf{Domain Rule Violation} occurs when an agent's action violates domain rules of the workload. This failure type is the most common in the airline workload, accounting for 21\%, 42\%, and 50\% of all failures under GPT 4.1, GPT 4.1 mini, and o3, respectively. 
The domain rule most commonly violated is the modification policy of the airline, which restricts the modification of reservations based on the time of purchase, flight cabin, and user membership.
Similar to tool output processing failures, we observe that while the LLM can correctly apply domain rules when presented in isolation, the addition of extra information makes it more difficult for the agent to adhere to domain rules correctly.

In our analysis, we further observe two subtypes of domain rule violations. The first is \textit{invalid action}, where the agent performs an action that violates a domain rule (e.g., modifying a flight that is not allowed to be modified). The second type is a \textit{lack of correct action}, where the agent incorrectly concludes that a valid action is forbidden (e.g., informing a user they cannot upgrade an eligible ticket). 
Addressing these two types of domain rule violations requires different optimizations, as we will detail in Section \ref{subsec:opt2}. 

\textbf{User Instruction Following Failure } occurs when the agent fails to follow specific instructions from the user. For example, when asked to create a file with specific content, an agent in the file system workload created the file but populated it with its own interpretation of the content. This failure was most frequent in the file system workload, constituting 40\%, 40\%, and 89\% of its failures under GPT 4.1, GPT 4.1 mini, and o3. 
Environment optimizations do not directly address this type of failure since we do not modify the user instructions. We still include this for completeness of our taxonomy to account for all agent failures we observe and to understand the limitations of system-level optimization.

\begin{figure}
  \centering
    \includegraphics[width=1\columnwidth]{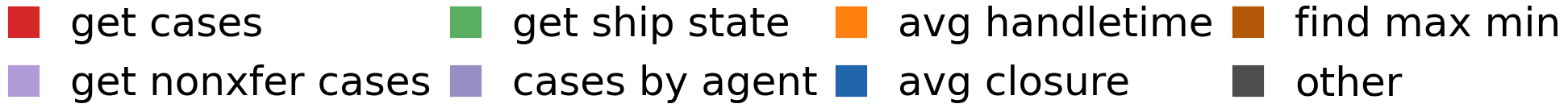}
  \hspace{0.1mm}
  \vspace{-6mm}
  
    \subfloat[CRM tasks baseline.]{
\includegraphics[width=0.5\columnwidth]{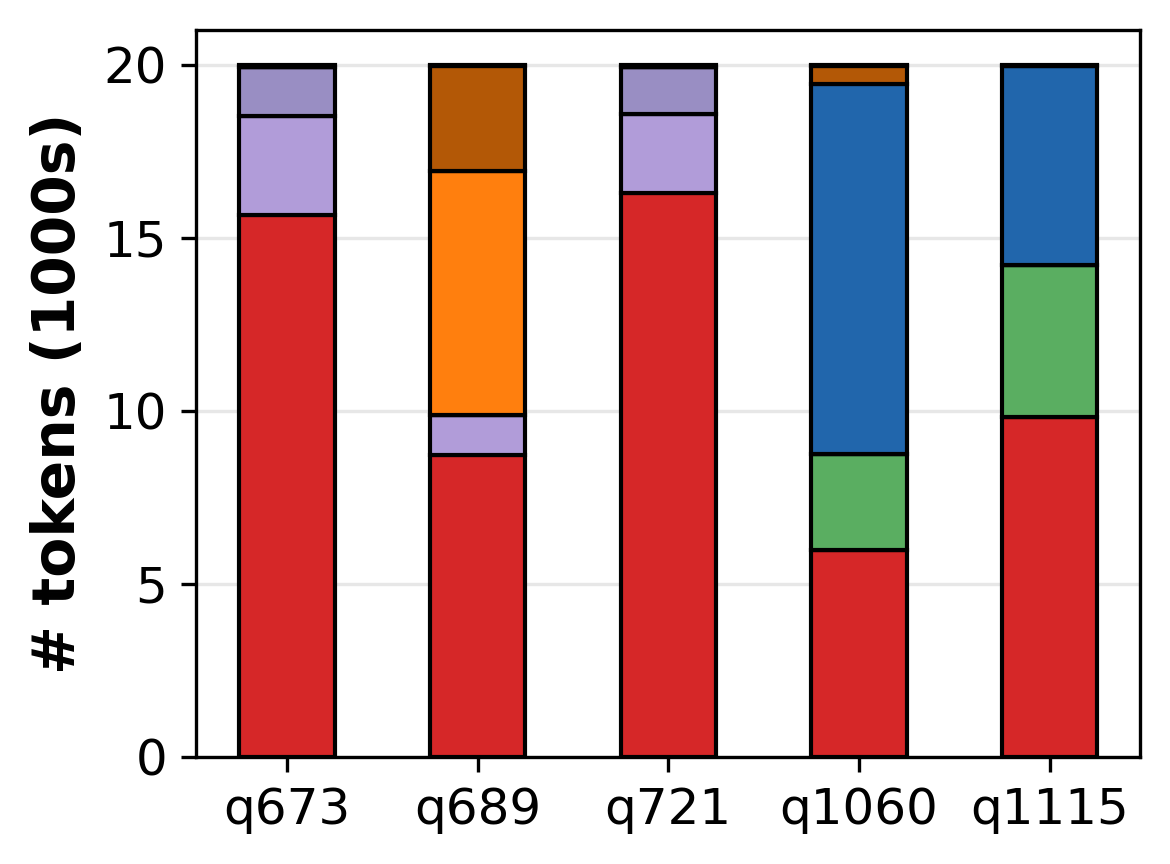}}
  \subfloat[CRM tasks with speculative agentic actions.]{
\includegraphics[width=0.5\columnwidth]{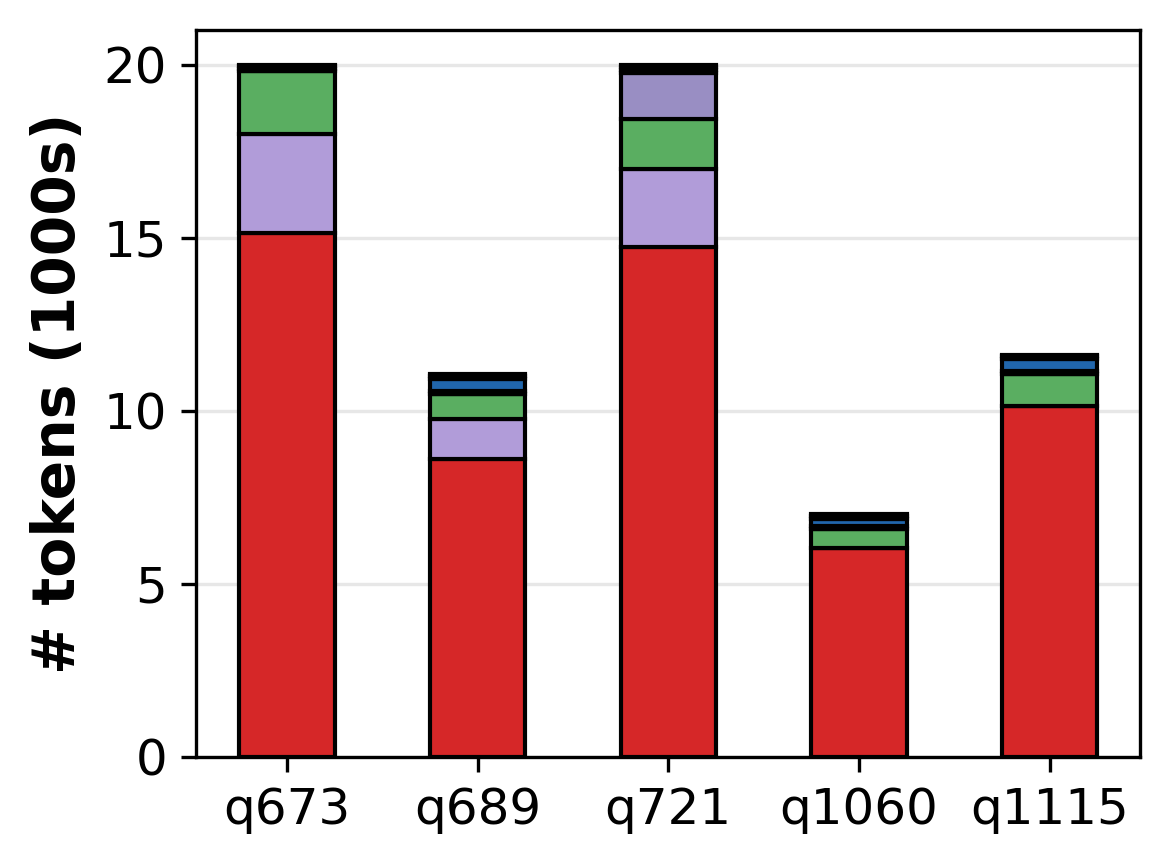}}

\vspace{3mm}
    \includegraphics[width=0.68\columnwidth]{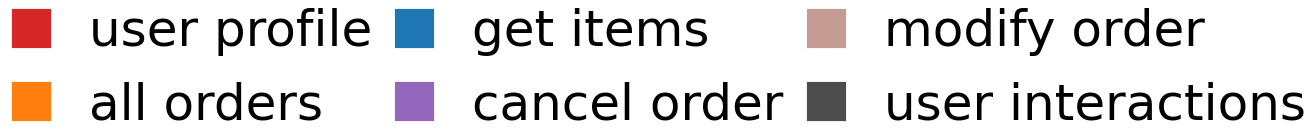}
  \hspace{0.1mm}
  \vspace{-7mm}
  
  \hspace{0.3mm}
    \subfloat[Retail tasks baseline.]{
\includegraphics[width=0.5\columnwidth]{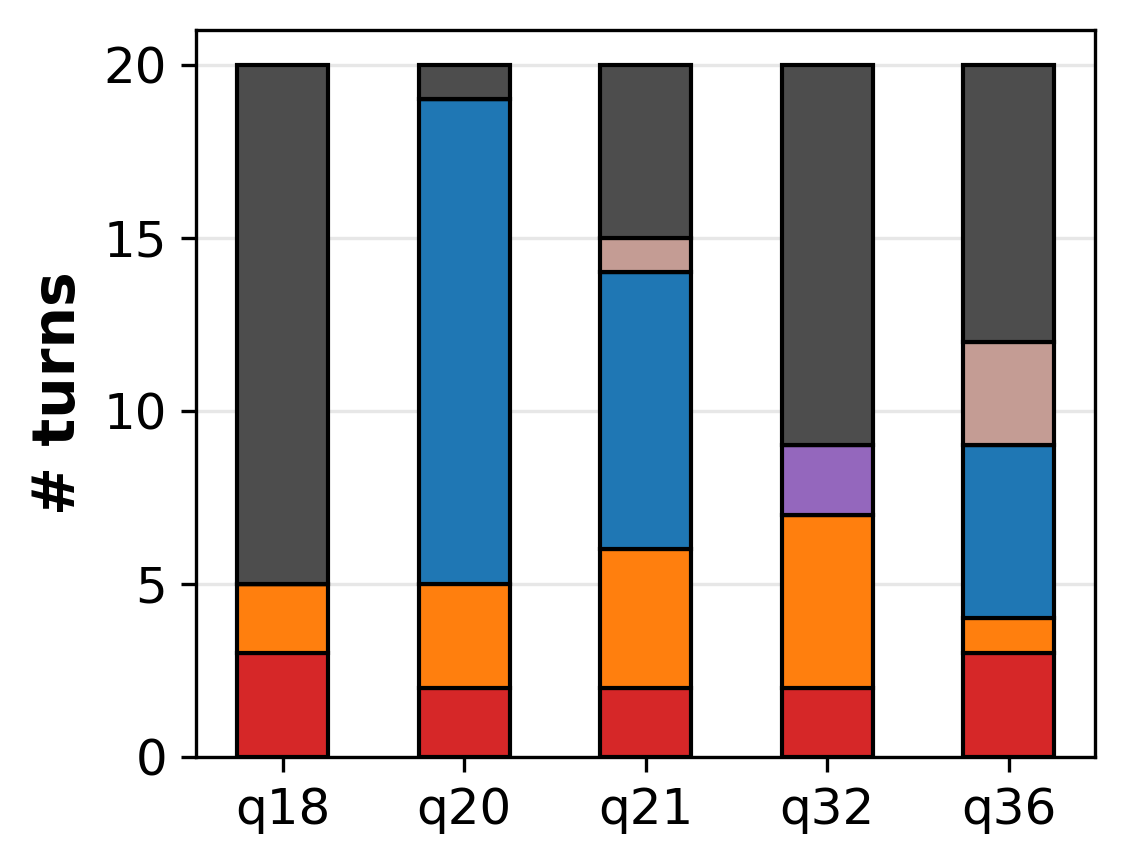}}
  \subfloat[Retail tasks with speculative agentic actions.]{
\includegraphics[width=0.5\columnwidth]{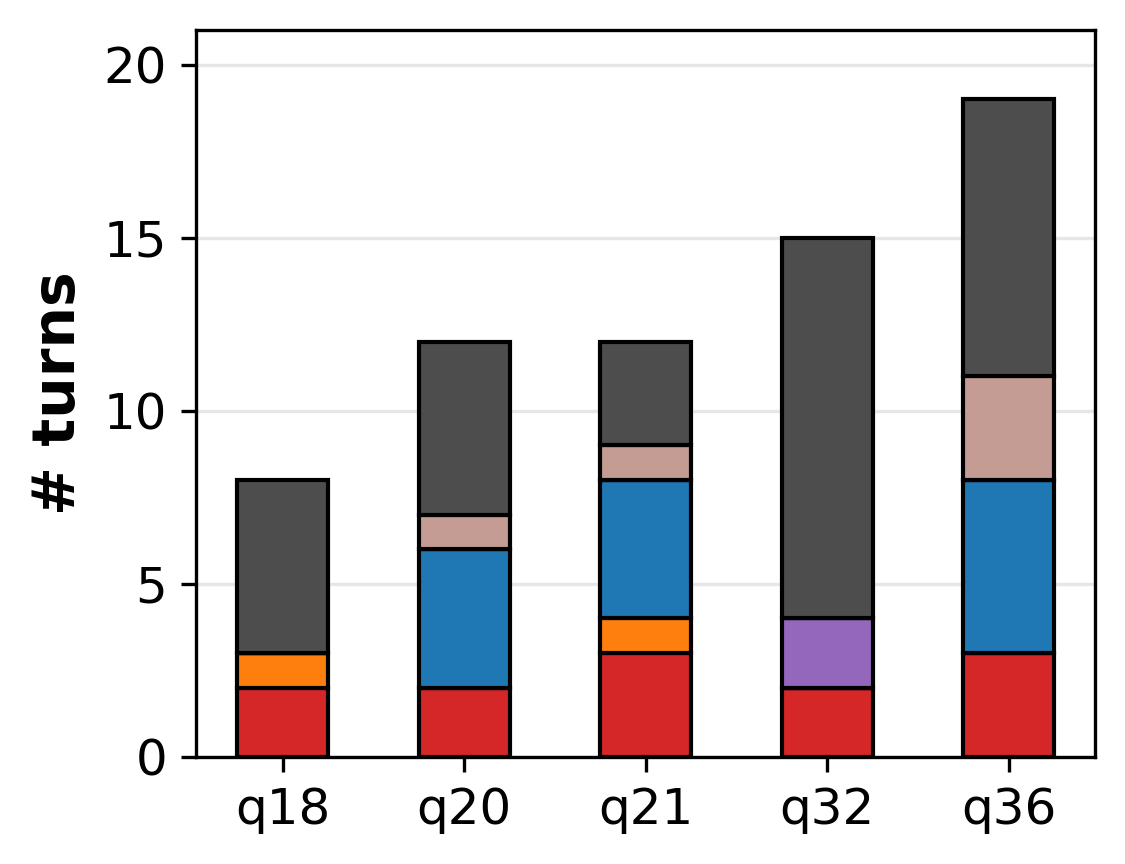}}
  \caption{Token/turn breakdown on 
  all GPT 4.1 tasks failed due to resource exhaustion on CRM and retail. Each bar represents a task failed due to resource exhaustion. Stacks show token/turn breakdown by subtasks.}
  \Description{}
  \label{fig:token_turn_distr_after}
\end{figure}

\subsection{Resource Exhaustion Failure} \label{subsec:turn_limit_failures}

This failure occurs when an agent cannot complete a task before reaching the maximum number of turns or tokens allocated. 
Such limits are standard practice in agent benchmarks to control cost, bound runtime, and penalize inefficient reasoning~\cite{taubench, bfcl_v3, medagentbench, crmarena, agentbench, toolsandbox}. We set a limit of 20 turns and 20,000 tokens per task, consistent with prior work~\cite{agentbench, bfcl, bfcl_v3, crmarena}, and have verified that all tasks are solvable within these constraints.
As shown in Figure~\ref{fig:failure_distr}, resource exhaustion is a major source of failure in the retail and CRM workloads, accounting for 45\% and 83\% of failures respectively.
Token exhaustion is the primary issue in CRM. As detailed in Figure~\ref{fig:token_turn_distr_after}a, subtasks such as \texttt{get\_cases} can return large volumes of data (e.g., over 200 cases in task 721). Subsequent tools that require this data as input argument, such as \texttt{calc\_avg}, force the agent to generate a correspondingly large argument list, which further exhausts the token budget.
On the other hand, turn exhaustion is the dominant issue in retail. As shown in Figure~\ref{fig:token_turn_distr_after}b, agents often exceed the turn limit through lengthy \texttt{user interactions} or repeated calls to \texttt{get\_items} to fetch data. 
We also observe that the agent consistently spends 4-7 turns to retrieve the user profile, then all user orders.
These examples show that inefficient interactions, whether from verbose tool outputs or repetitive queries, are a primary cause of resource exhaustion.


\begin{figure}
  \centering
  \includegraphics[width=1.0\columnwidth]{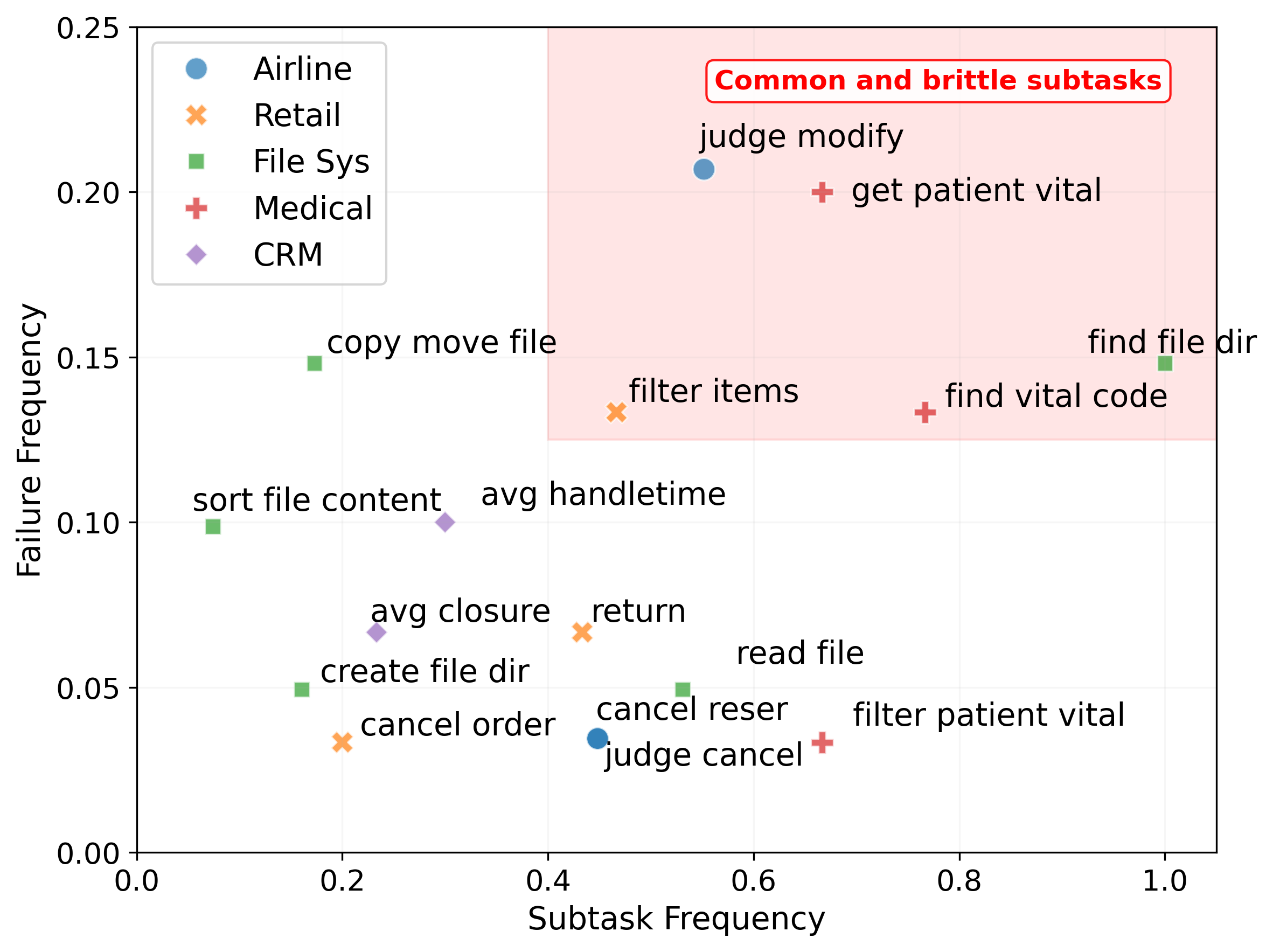}
  \caption{Subtask frequency vs. failure frequency (GPT 4.1).}
  \Description{}
  \label{fig:subtask_freq_vs_failure_freq}
\end{figure}

\subsection{Subtask Failure Analysis}
We next link the failure categories back to the specific subtasks where they arise. For every subtask, we compute 1) failure rate: the number of times the subtask failed divided by the number of times the subtask appears 2) frequency: the number of times the subtask appears divided by the total number of subtasks.
Figure~\ref{fig:subtask_freq_vs_failure_freq} plots failure rate (y-axis) against frequency (x-axis) for all workloads. Subtasks in the upper-right quadrant are both common and brittle, and therefore represent the highest-leverage optimization targets.
Across workloads, we identify the following brittle subtasks, which are the primary candidates for environment improvements.
\begin{CompactItemize}
  \item Airline: \texttt{judge reservation modification}
  \item Retail: \texttt{filter product items}
  \item File system: \texttt{find file/directory}
  \item Medical: \texttt{find vital code}, \texttt{get patient vital data}
  \item CRM: \texttt{compute average handle time}
\end{CompactItemize}


\section{System Environment Optimizations}  \label{sec:opts}
In this section, we utilize insights from Section \ref{sec:workload_analysis} and our failure taxonomy to optimize the system environment.
Our optimization focuses exclusively on the agent's environment. We do not alter the user's task, the agent's core logic (i.e., the underlying model and prompting techniques), or the task's correctness criteria. We do not add new tools, only augment existing ones.
We propose \projectname{}, a set of targeted optimizations corresponding to each failure category, as shown in Table \ref{tab:failure_taxonomy}.
We now describe each optimization in detail.

\subsection{Enhancing Environment Observability} \label{subsec:opt1}
To address exploration failures, our principle is to improve the agent's observability of the environment.

\begin{figure}
  \centering
  \includegraphics[width=\columnwidth]{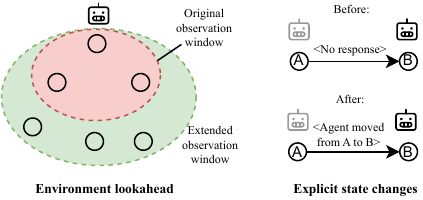}
  \caption{Illustration of exploration optimizations.}
  \Description{}
  \label{fig:explore_opt_illustrate}
\end{figure}

\textbf{Environment Lookahead. }
In Section \ref{subsec:exploration_failures}, we observe that an agent's exploration is often inefficient because the critical context lies beyond its observation window. The agent then explores the environment in a best-effort manner, hoping to reach the desired information by chance. 
To address this type of failure, we introduce \textit{environment lookahead}, a technique where exploration tools are augmented to provide a "peek" into adjacent locations in the environment with respect to the agent's current position. As illustrated in Figure~\ref{fig:explore_opt_illustrate}, this expanded observation window gives the agent a broader view of its surroundings, increasing the probability of successful navigation. 
For instance, when the environment detects that the agent is retrieving paginated patient records, it can return a message indicating the total number of records and the number of records already retrieved. This hint effectively guides the agent towards the correct exploration actions and reduces the chance of missing critical context.

\textbf{Explicit State Changes. }
This optimization targets state awareness failures, which occur when an agent's internal understanding of its current location within the environment is incorrect.
This failure causes the agent to issue subsequent commands from an incorrect assumption about its current state.
To mitigate this, we augment the environment to explicitly communicate the agent's state changes. Whenever a tool call modifies the agent's state, we augment the tool's response to include a confirmation of the new state, as shown in Figure~\ref{fig:explore_opt_illustrate}. 
For example, the response from the \texttt{cd} command is augmented to confirm the new working directory and list its contents, directly reinforcing the agent's understanding of its location.
From the agent's perspective, this optimization turns a reasoning task (tracking the current state through a potentially complex sequence of exploration) into a simple retrieval task (reading the state from the last tool response).

\subsection{Offloading Common Computational Patterns} \label{subsec:opt2}
In Section \ref{subsec:exploitation_failures}, we observe that while LLMs are fundamentally capable of performing complex reasoning tasks, they still fail to perform simple mathematical operations and rule-based reasoning in agentic workloads due to context distractions.
To address exploitation failures, our key principle is to \textit{minimize the amount of reasoning} the agent has to perform to reduce the chance of information synthesis failures.
To do this, we offload common computations/reasoning operations from the agent to the environment.
Unlike the agent, we can perform operations such as sorting, calculation, and constraint checking reliably in the environment.

\textbf{Offload Tool Output Processing. }
This optimization targets tool output processing failures. We offload common computations such as sorting (Section \ref{subsec:exploitation_failures}) by augmenting tools to return pre-computed results alongside their primary output. For instance, a tool that retrieves products for the retail agent is augmented to also return the products sorted by price.
%
We note that this optimization is opportunistic, as the agent may not always need the pre-computed results. 
However, since such computations are common, the benefits of having pre-computed results readily available outweigh the minor cost of slightly longer tool responses\footnote{The alternative, creating dedicated tools (e.g., \texttt{get\_cheapest\_item()}), would increase overall cost and latency due to an additional tool-use turn.}.

\textbf{Offload Domain Rule Validation. }
To address domain rule violations, we offload rule validation from the agent to the environment. 
Our analysis in Section~\ref{subsec:exploitation_failures} identified two subtypes of these violations: performing invalid action and lack of correct action.
We address invalid actions by implementing \textit{environment guardrails}, which embed rule validation directly into tools. 
For instance, when the airline agent attempts to modify a reservation, the tool first validates the request against domain rules (e.g., time of purchase, flight cabin) and rejects the action if a rule is violated.
For failures caused by a lack of correct action, guardrails are ineffective because the agent makes no attempt. 
Instead, we enrich tool outputs with \textit{domain rule hints}. 
For example, when the agent retrieves the user's reservation information, the tool response is augmented to indicate whether each reservation is modifiable based on domain rules. 
This optimization transforms the agent's task from complex rule deduction into simple information retrieval, significantly reducing such errors.

\begin{figure}[H]
  \centering
  \includegraphics[width=\columnwidth]{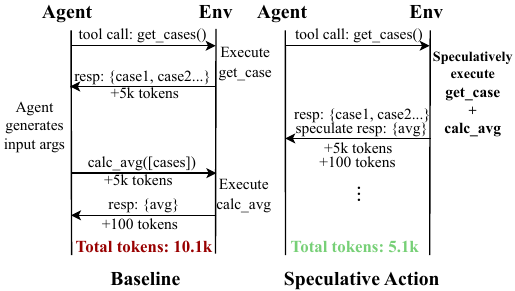}
  \caption{Speculative agentic actions example (CRM task).}
  \Description{}
  \label{fig:spec_act_illustration}
\end{figure}

\subsection{Speculative Agentic Actions} \label{subsec:opt3}
This optimization targets resource exhaustion failures
 by speculatively executing likely subtask sequences in a single turn. 
Using our subtask transition analysis, when an agent calls a tool that typically precedes another, the environment preemptively executes both and bundles the results into a single response.
We present an example in Figure \ref{fig:spec_act_illustration}. 
As we observed in section \ref{subsec:turn_limit_failures}, 
in the CRM workload, \texttt{get\_cases} often returns a large set of results.
The subsequent \texttt{calc\_avg} function call requires the list of cases as an input argument, requiring the agent to generate a substantial number of additional tokens. 
With speculation, when the agent invokes \texttt{get\_cases}, the environment speculatively also executes \texttt{calc\_avg} using the output from \texttt{get\_cases}. 
The response from \texttt{calc\_avg} is bundled with \texttt{get\_cases} output and returned to the agent.
When this speculation matches the agent’s next step (a “hit”), we save the tokens required to generate the input argument of \texttt{calc\_avg}.

Speculative agentic actions also reduce turn count. In the same CRM example, a speculation hit eliminates an additional agent-environment round trip, as the average result is already available. 
Thus, we apply speculative agentic actions in retail workload to address tasks where the turn limit is reached.
In the retail workload, we observe that agents almost always retrieve the user profile before fetching the user’s orders (Figure~\ref{fig:merged_trees}). Thus, we execute \texttt{get\_user\_orders} within \texttt{get\_user\_profile} speculatively.

We note that this optimization is probabilistic. In situations where, e.g., the CRM agent calls \texttt{get\_cases} without intending to execute \texttt{calc\_avg}, the speculation results in a “miss,” and the extra tokens consumed by \texttt{calc\_avg} output are wasted  (the ``miss cost'').
As with other speculative techniques (e.g., CPU cache prefetching), this optimization is most effective when the miss cost is low.

\section{Evaluation} \label{sec:evaluation}

\subsection{Success Rate}

\begin{figure}[t]
  \centering
  \includegraphics[width=1\columnwidth]{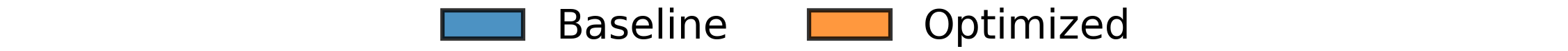}
  \hspace{0.1mm}
  \vspace{-7mm}
  
  \subfloat[GPT 4.1 Analysis set.]{
  \includegraphics[width=0.5\linewidth]{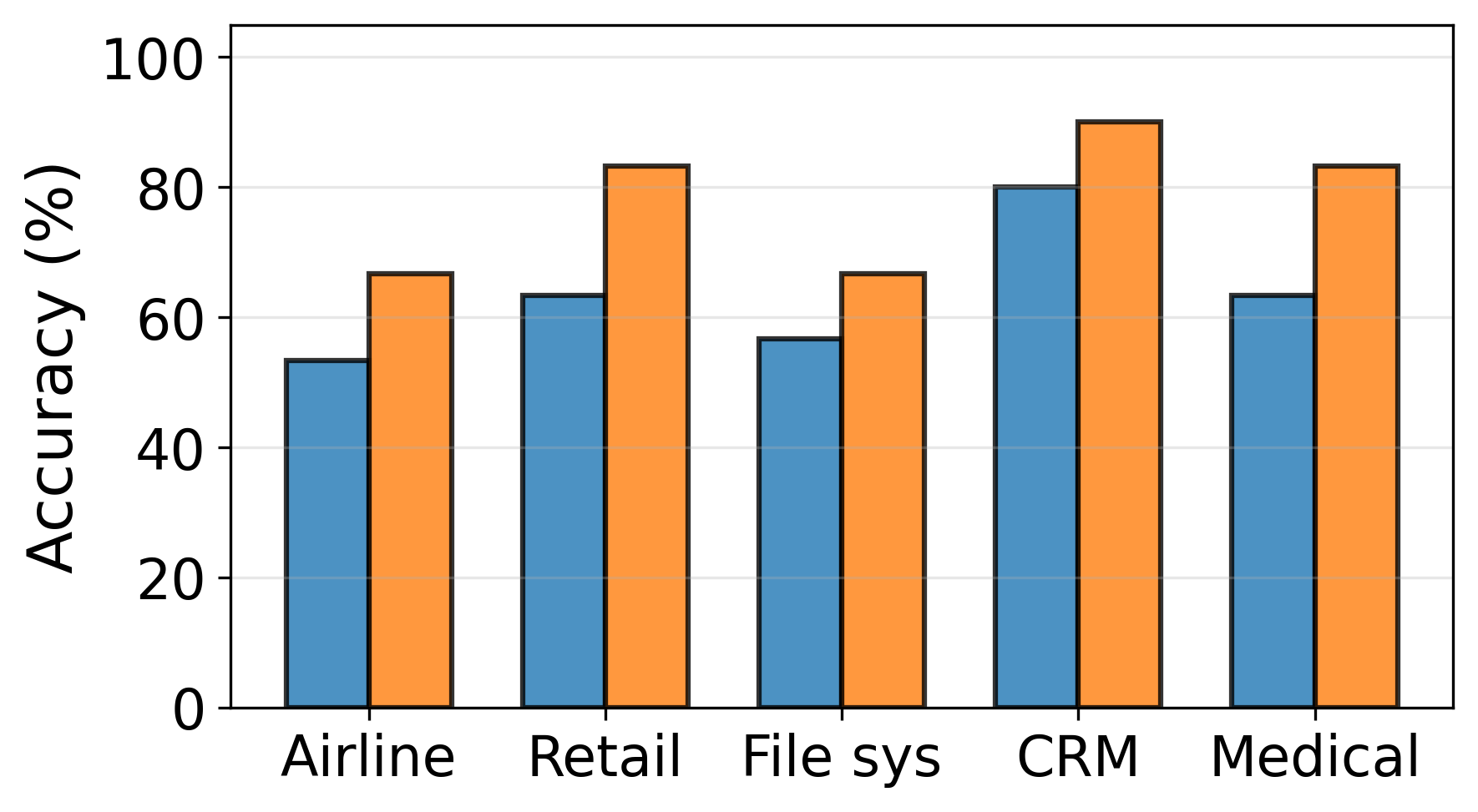}}
  \subfloat[GPT 4.1 Evaluation set.]{
  \includegraphics[width=0.5\linewidth]{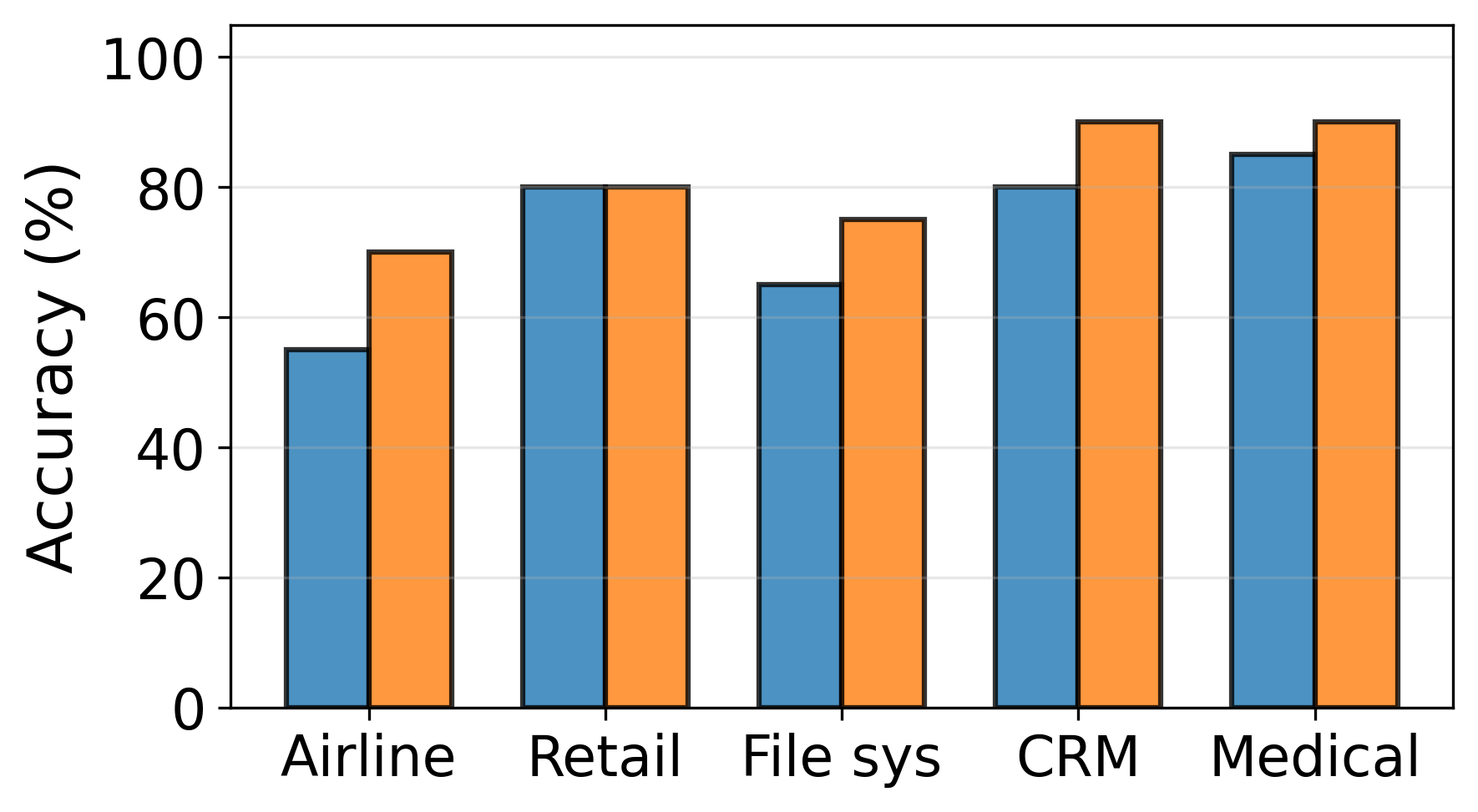}
  }
  \vspace{-3mm}
  
   \subfloat[GPT 4.1 mini Analysis set.]{
   \includegraphics[width=0.5\linewidth]{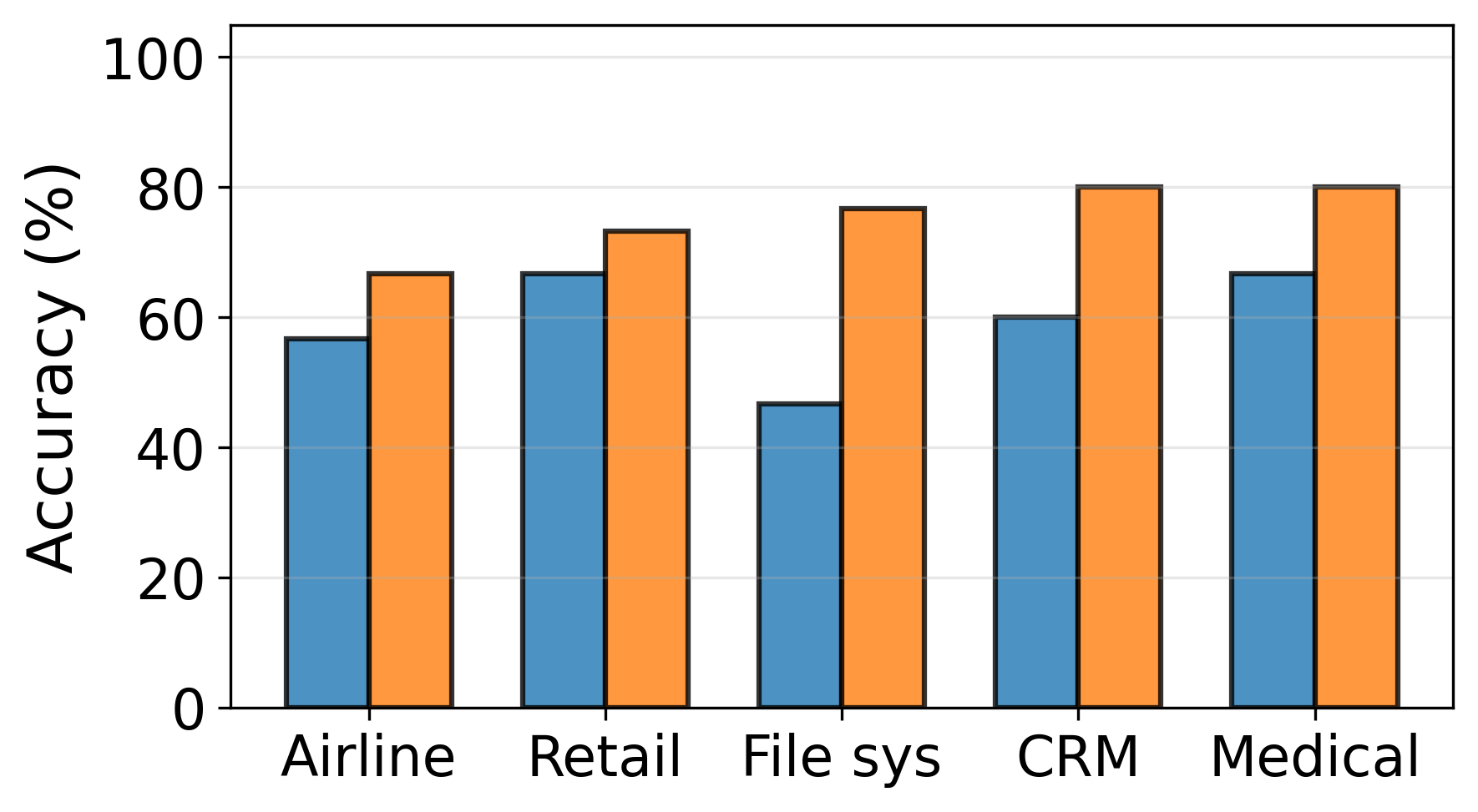}}
   \subfloat[GPT 4.1 mini Evaluation set.]{
   \includegraphics[width=0.5\linewidth]{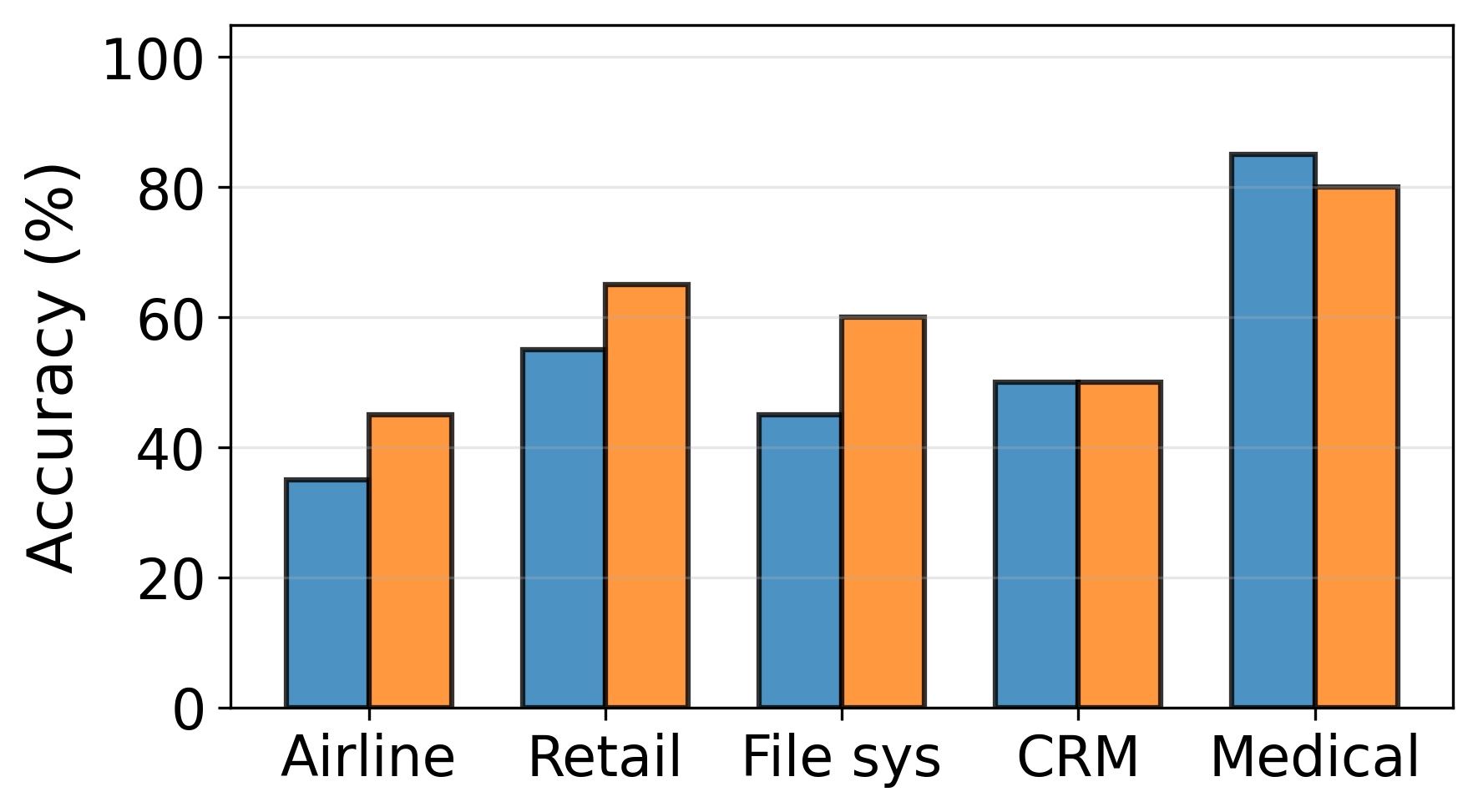}
   }
  \vspace{-3mm}
   \subfloat[o3 Analysis set.]{
   \includegraphics[width=0.5\linewidth]{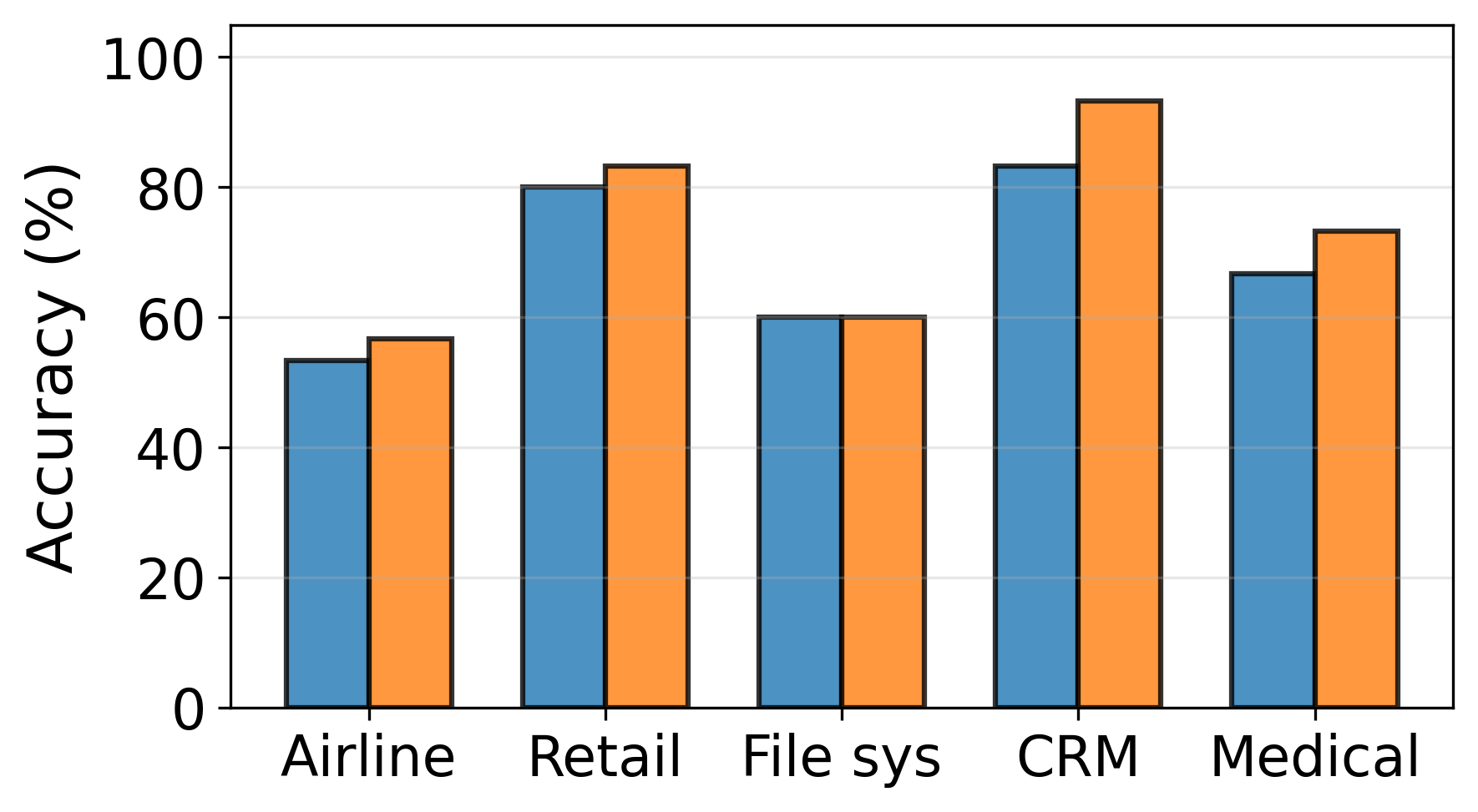}}
    \subfloat[o3 Evaluation set.]{
   \includegraphics[width=0.5\linewidth]{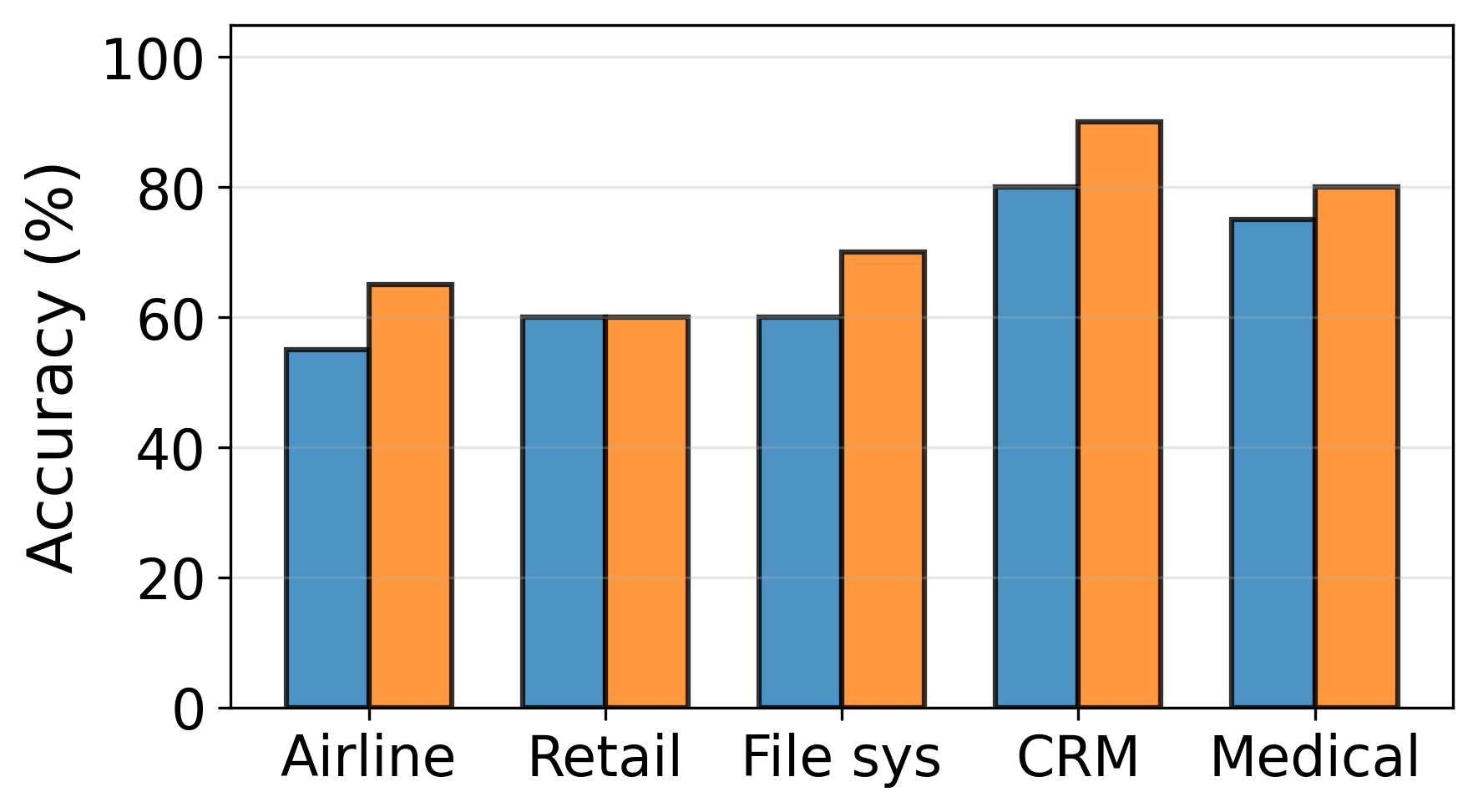}
   }
  \caption{Overall task success rates.}
\label{fig:overall_acc}
\Description{}
\end{figure}

Figure~\ref{fig:overall_acc} shows the success rate before (Baseline) and after 
(Optimized) system environment optimizations. We evaluate three models representing different capability levels: general-purpose GPT 4.1, the smaller, economical GPT 4.1 mini, and the state-of-the-art reasoning model o3. We show the results for the analysis set and evaluation set separately.

On average, the environment optimizations lead to 10.3\%, 6.7\%, 12.5\%, 10.0\%, and 
7.5\% improvement in success rate for the five workloads, respectively.
To contextualize these gains, the improvements from our system-level approach are comparable to or greater than those typically seen between major LLM model generations, which are often a few percent \cite{o3, claude4, gpt41}. For instance, the reported success rate improvement between o1 and o3 on two of our workloads, airline and retail, was 2-3\%~\cite{o3}. This result demonstrates that system-for-agent optimization is a powerful and complementary path to enhancing agent reliability.
On average, our environment optimizations improve the success rate of the analysis set by 11.8\% and the evaluation set by 7.0\%. The higher improvement on the analysis set is expected, as our methodology directly targets failures observed within it. The improvement on the unseen evaluation set demonstrates that our optimizations generalize effectively to new tasks.

\subsection{Failure Breakdown} \label{subsec:failure_breakdown}
We analyze the distribution of failure types before and after environment 
optimizations in Table~\ref{tab:failure_breakdown_gpt41}. We observe that our targeted optimizations successfully reduce the most frequent failures. For instance, the compute offloading optimization cuts \textit{Tool Output Processing} failures in the airline workload by 60\% (from 5 to 2). Similarly, exploration optimizations eliminated the majority of exploration failures in airline, file system, and medical workload.

A key observation is that mitigating one type of failure can reveal underlying, previously masked failures. For instance, after addressing navigation failures in the medical workload, agents gathered more extensive patient data, which in turn caused some tasks to exceed the token limit, triggering new \textit{Resource Exhaustion Failure} failures. A similar effect occurred in the airline and retail workloads, where resolving exploitation failures revealed \textit{User Instruction Following} failures not previously observed. This observation highlights the compounding nature of agent failures and the need for a holistic approach to improving agent success rate.

\begin{table}[t]
    \centering
    \small
    \setlength{\tabcolsep}{1pt}
    \caption{Failure breakdown (counts) before (Base) and after (Opt) system optimizations for GPT~4.1 in the analysis set (30 tasks total in each workload). Dashed lines represent 0 failures before and after optimizations. Underscores show cases where failures have been reduced.}
    \resizebox{\linewidth}{!}{%
    \begin{tabular}{l cc cc cc cc cc}
      \toprule
      & \multicolumn{2}{c}{Airline} & \multicolumn{2}{c}{Retail} & \multicolumn{2}{c}{File Sys} & \multicolumn{2}{c}{CRM} & \multicolumn{2}{c}{Medical} \\
      \cmidrule(lr){2-3} \cmidrule(lr){4-5} \cmidrule(lr){6-7} \cmidrule(lr){8-9} \cmidrule(lr){10-11}
      \textbf{Failure category} & Base & Opt & Base & Opt & Base & Opt & Base & Opt & Base & Opt \\
      \midrule
      State Awareness & - & - & - & - & 3 & \underline{1} & - & - & - & - \\
      State-space Navigation & 2 & \underline{0} & 0 & 1 & 1 & 1 & - & - & 10 & \underline{0} \\
      Domain Rule Violation & 3 & 3 & 1 & \underline{0} & - & - & - & - & 0 & 1 \\
      Tool Output Processing & 5 & \underline{2} & 5 & \underline{1} & 2 & 2 & 1 & 1 & 1 & \underline{0} \\
      User Inst. Following & 0 & 2 & 0 & 3 & 7 & \underline{6} & - & - & - & - \\
      Resource Exhaustion & 2 & 2 & 5 & \underline{0} & - & - & 5 & \underline{2} & 0 & 4 \\
      \bottomrule
    \end{tabular}%
    }
    \label{tab:failure_breakdown_gpt41}
    \Description{}
  \end{table}
  
\subsubsection{Speculative Agentic Actions} \label{subsec:spec_exec_eval}
Figures \ref{fig:token_turn_distr_after}c and d show the effects of speculative agentic actions on the same set of tasks analyzed in Section \ref{subsec:turn_limit_failures}.
Out of the 5 CRM tasks, Q689, Q1060, and Q1115 become correct by speculatively executing \texttt{calculate\_avg} after \texttt{get\_cases}, reducing the token cost of \texttt{avg handletime} and \texttt{avg closure} as described in Section \ref{subsec:opt3}
Q673 and Q721 remain unsolved as the \texttt{get\_cases} function returns a large amount of data that still exceeds the token limit despite speculative action savings.
%
Among the 5 retail tasks, Q21, Q32, and Q36 become correct. 
The turn savings come from speculatively retrieving all user orders within the \texttt{get\_user\_profile} function.

  

  

\subsection{Cost Analysis}
\begin{figure}[t]
  \centering
  \subfloat[Monetary cost breakdown.]{
  \includegraphics[width=0.50\columnwidth]{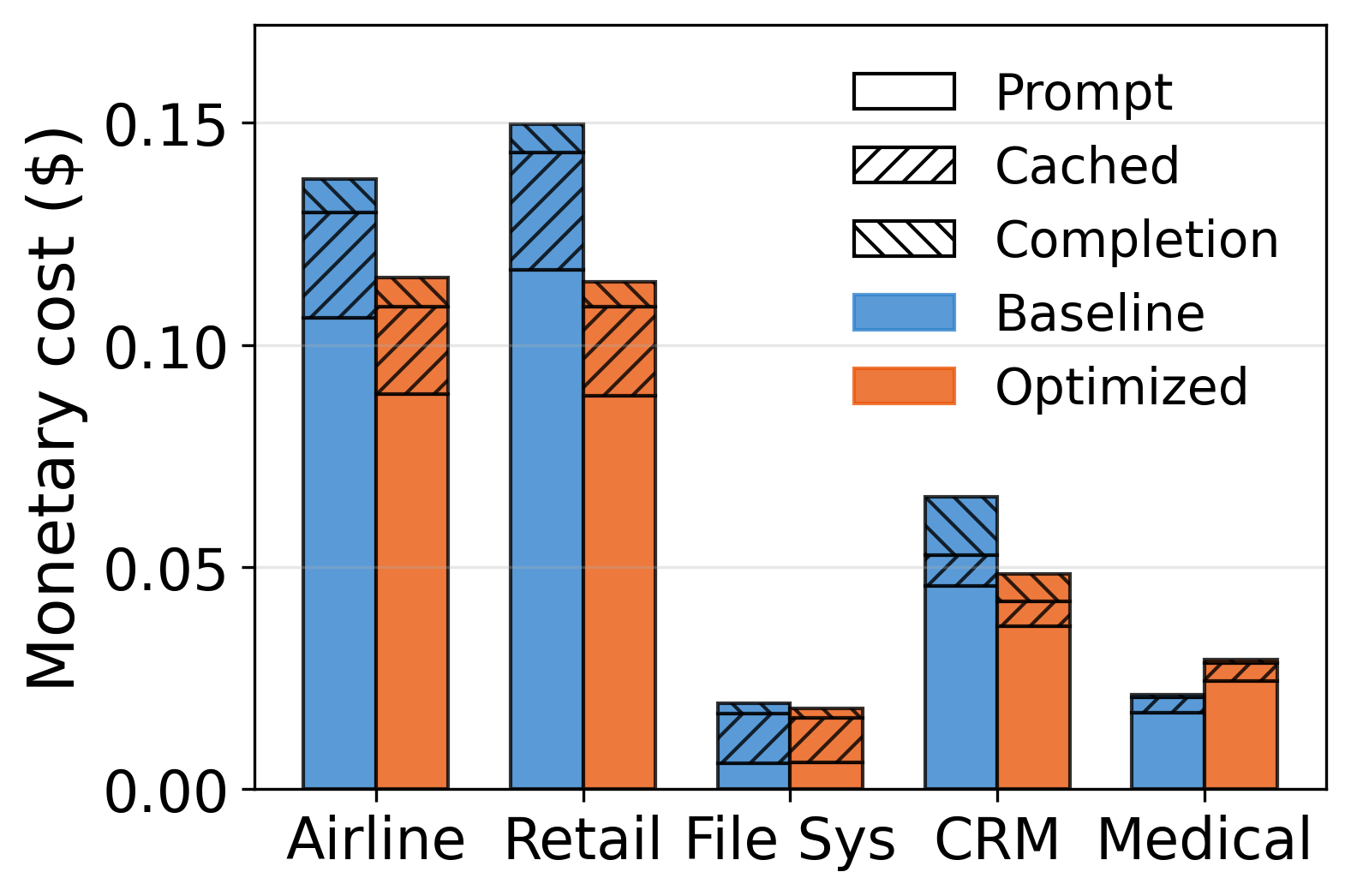}}
  \subfloat[Turns breakdown]{
  \includegraphics[width=0.50\columnwidth]{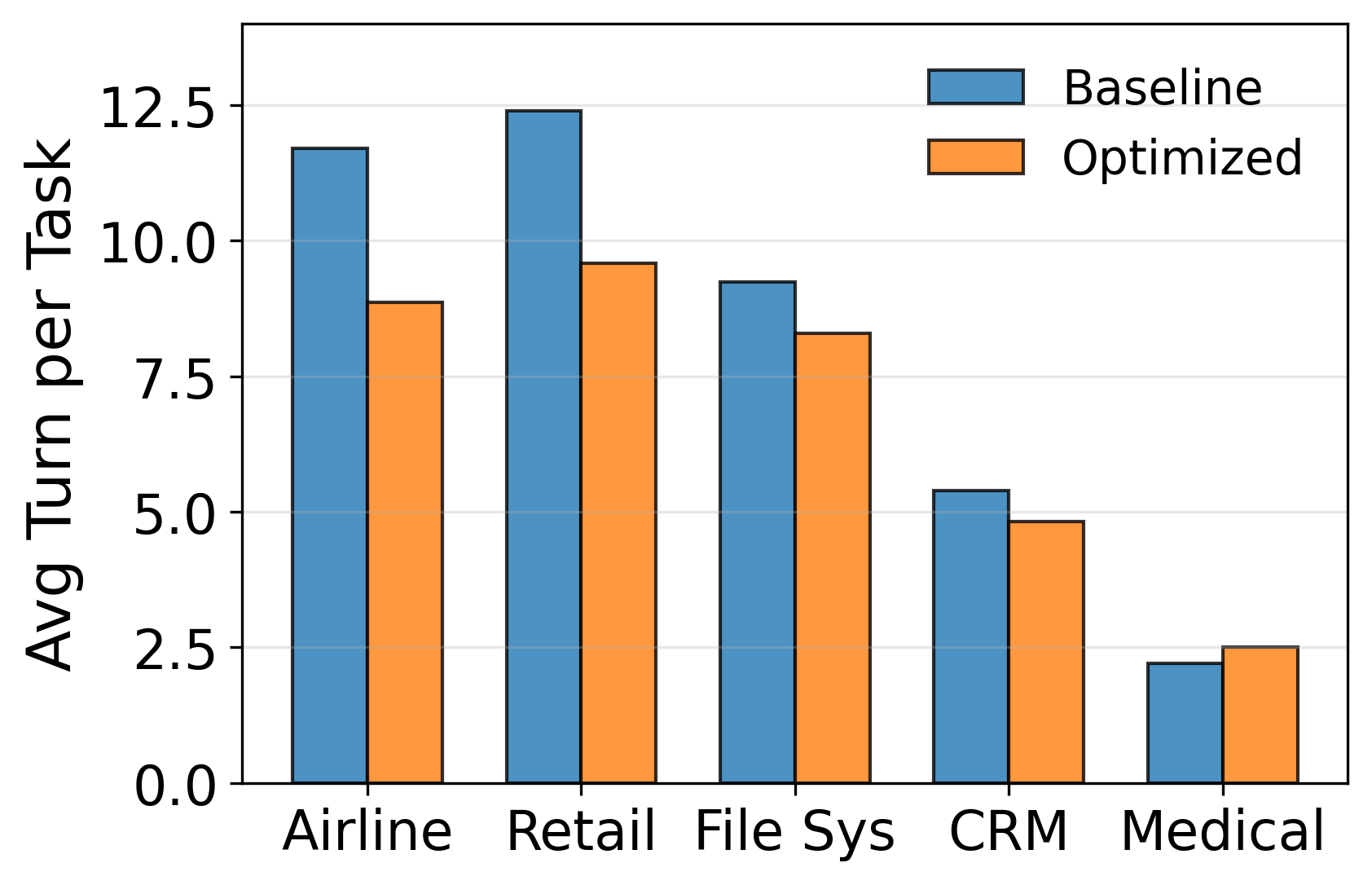}
  }
  \caption{GPT 4.1 monetary cost and turns before and after environment optimizations.}
\label{fig:overall_cost}
\Description{}
\end{figure}

Environment optimizations reduce cost in addition to improving accuracy.
Figure \ref{fig:overall_cost} reports the monetary cost and turns before and after our optimizations.
We compute cost using OpenAI API pricing and decompose it into three categories according to the standard LLM inference pricing model: ``Prompt'' (input tokens that miss the prompt cache), ``Cached'' (input tokens that hit the prompt cache), and ``Completion'' (output tokens, including reasoning tokens for o3).
On average, environment optimizations reduce monetary cost by 17.4\%, 7.1\%, and 17.7\% for GPT 4.1, GPT 4.1 mini, and o3, respectively. 
These cost savings are driven by the reduction in the number of turns consumed to complete the task: 16.8\%, 7.2\%, and 16.6\% for the same models. Turn reductions are a result of speculative agentic actions, as we demonstrated in Section \ref{subsec:spec_exec_eval}.

One exception is the medical workload, where cost increases by 29-33\% after optimization.
This is because in the baseline environment, agents often fail due to insufficient exploration (Section \ref{sec:agent_failures}), terminating early without retrieving all relevant data.
The observability enhancement enables the agent to correctly gather more information, despite higher token usage. We consider this a favorable tradeoff.

\subsection{Fine-tuning}
\begin{figure}[t]
  \centering
  \subfloat[Analysis set.]{
  \includegraphics[width=0.48\linewidth]{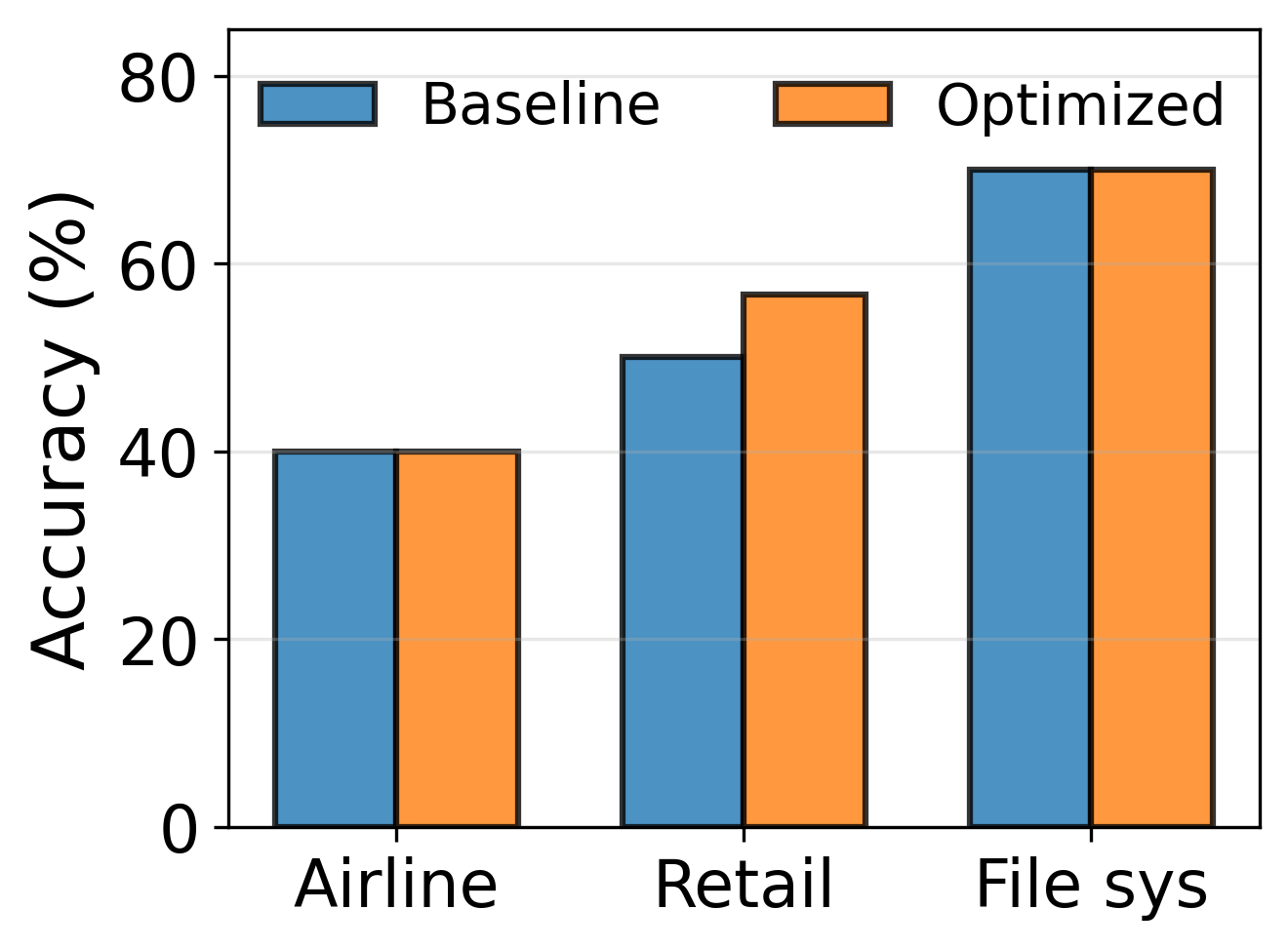}}
  \hspace{-2mm}
  \hfill
  \subfloat[Evaluation set.]{
  \includegraphics[width=0.48\linewidth]{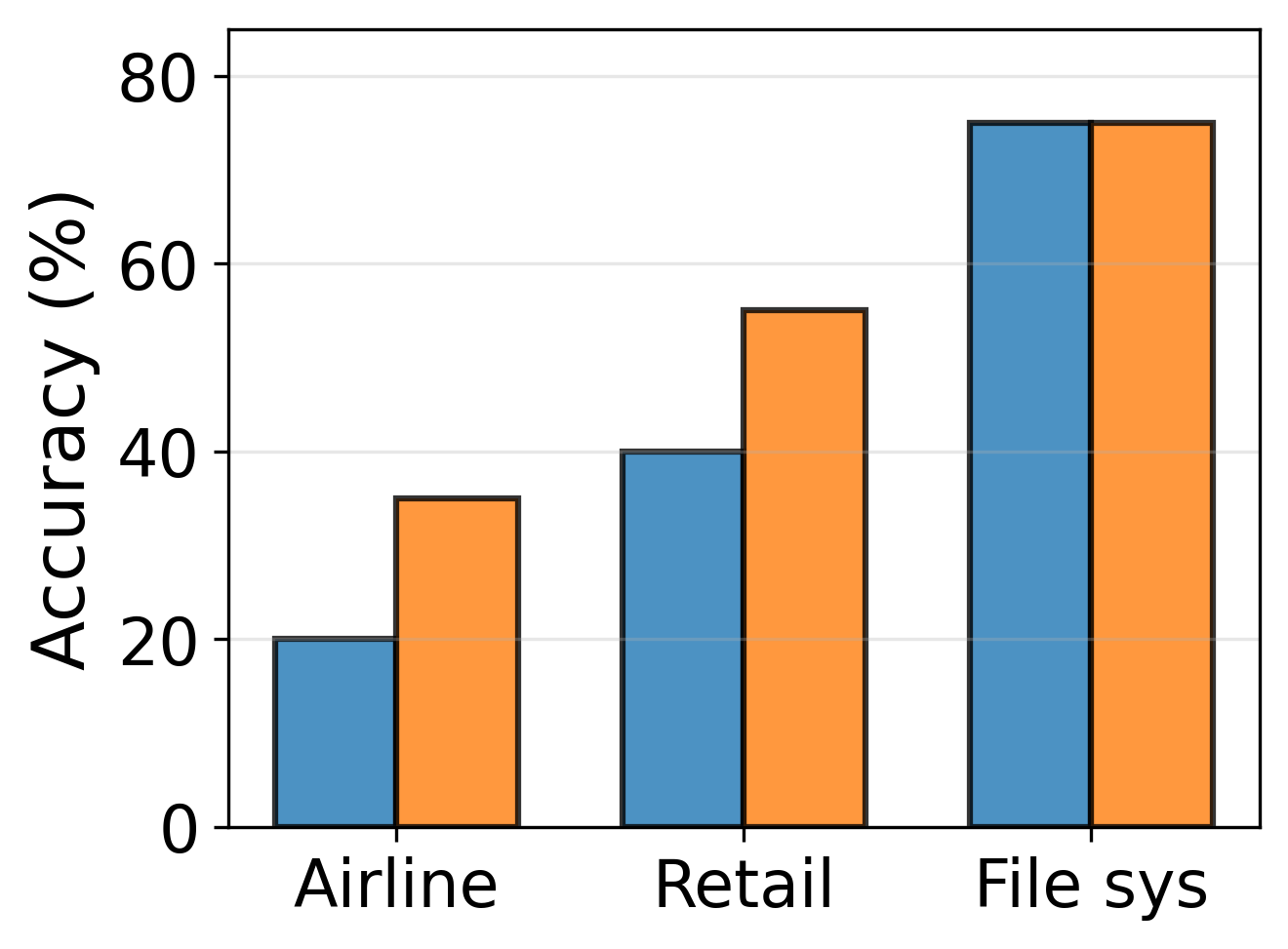}
  }
  \caption{Task success rates on xLAM-2 8B LLM model.}
\label{fig:xlam_acc}
\Description{}
\end{figure}

To evaluate the effect of our optimizations on fine-tuned models, we test them on xLAM-2 8B, a model fine-tuned from Llama 3 8B \cite{llama3} that achieves state-of-the-art accuracy compared to models of similar size on $\tau$-bench (airline and retail) and BFCL (file system) workloads \cite{xlam3, bfcl_v3}. 
We deploy this open-source model locally using an NVIDIA RTX A6000 GPU and evaluate its performance, shown in Figure \ref{fig:xlam_acc}
We observe that environment optimizations improve the average task success rate by 3.3\% on the analysis set and 8.8\% on the evaluation set. This result demonstrates that system-level optimizations are complementary to model fine-tuning; they can further enhance agent reliability beyond the benefits gained from fine-tuning.
We notice that xLAM-2 8B performs exceptionally well on file system, and our optimizations provide no additional improvement. 
By analyzing the failed tasks in this workload, we find that most failures remaining are due to user instruction following failures, which cannot be directly addressed by environment optimizations.
This is consistent with our findings for other models in Figure \ref{fig:failure_distr}.

The key advantage of system-for-agent is its lower development effort. Although potentially effective, fine-tuning is a resource-intensive and error-prone process \cite{ragvsfinetune, databricksfinetune, coherefinetune}. It involves a complex, multi-stage pipeline of data preparation, model training, and evaluation. The tuned model is also prone to overfitting and catastrophic forgetting \cite{finetune_forget, finetune_forget2}.
In contrast, environment optimizations are lightweight and interpretable system-level modification that only involves modifying tool APIs, thus enabling rapid iterations. We therefore advocate that when building an agent-based system, developers should first prioritize optimizing the agent's environment. Fine-tuning should be considered only if these system-level improvements are insufficient.

\section{Related Works}

\paragraph{Failure Analysis of LLM Agents.}
Recent works systematically analyze agent failures through two main approaches. First,  taxonomies focusing on agent-centric failures (agent-for-system): Cemri et al. \cite{why_multiagent_fail} identify multi-agent failure modes such as incorrect inter-agent communication, while Deshpande et al. \cite{trail} categorize reasoning failures such as hallucinations and flawed planning. Our work complements this by focusing on failures due to agent-environment interaction (system-for-agent).
Another line of research focuses on automated failure localization methods. Zhang et al. \cite{which_agent_causes_failures} use LLMs to identify which agent step caused task failures, while Arabzadeh et al. \cite{agent_user_experience} introduce a framework for measuring agent misalignment with user expectations.  Our work complements these works by introducing a subtask abstraction for finer-grained failure decomposition and analysis.
In addition, unlike prior work that primarily identifies failures, we use our analysis to propose, implement, and empirically validate concrete environment optimizations that directly mitigate identified failure modes.

\paragraph{Specialized Agent Environments.}
One line of work shows that by designing specialized agentic environments, one can achieve superior performance. 
Zhang et. al. \cite{game_harness} observe that when an agent plays games, environment design can significantly impact agent performance. For example, whether the agent has access to screenshots of the gaming interface. 
Yang et. al. \cite{agentoccam} show that under web browsing tasks, by simply refining the agent's observation and action space, the agent's success rate can be improved. 
SWE-agent \cite{sweagent} designs custom agent-computer interface tools to improve the performance of software engineering agents. 
Our work is inspired by these works, so we try to find generalizable techniques for environment optimization.

A growing body of works shows that tailoring an agent's environment can significantly boost its accuracy. These works focus on creating highly specialized, domain-specific interfaces. For example, Zhang et al. \cite{game_harness} observe that an agent's success in games is highly dependent on the environment's design, such as whether the agent receives screenshots or only text. 
In the context of web browsing, Yang et al. \cite{agentoccam} show that refining the agent's observation and action space leads to substantial improvements in task success rates. 
SWE-agent \cite{sweagent} introduces custom agent-computer interface tools specifically designed to improve the performance of agents on software engineering tasks.
Our paper is inspired by these works. However, whereas prior efforts focus on developing specialized solutions for specific domains, our research aims to identify generalizable principles for environment optimization across various workloads.
\section{Conclusion}
In this paper, we present the system-for-agent design paradigm, which
treats the agent's environment as a first-class component in building more  reliable agentic systems.
We introduce a systematic methodology for analyzing agent-environment interactions, leading to a taxonomy of common failure modes. Based on this taxonomy, we develop \projectname{}, a set of targeted, system-level optimizations to mitigate these failures. Our evaluations demonstrate that these environment optimizations substantially improve agent task success rates across a variety of workloads and models.

\bibliographystyle{plain}
\bibliography{references}


\end{document}